\documentclass[peerreview,draftcls,onecolumn]{IEEEtran}

\usepackage[noadjust]{cite}

\ifCLASSINFOpdf
   \usepackage[pdftex]{graphicx}
\else
\fi
\usepackage[cmex10]{amsmath}
\usepackage{array}

\usepackage{mdwmath}
\usepackage{mdwtab}

\usepackage{eqparbox}

\usepackage{subfigure}

\usepackage{caption}
\usepackage{url}
 
\usepackage{stackrel}
\usepackage{etoolbox}
\usepackage{enumerate}
\usepackage{amssymb}
\usepackage{tikz, pgfplots}
\usetikzlibrary{calc}
\usetikzlibrary{arrows}
\usetikzlibrary{arrows,positioning} 
\usetikzlibrary{decorations.markings}
\usetikzlibrary{shapes}
\tikzset{
    >=stealth',
    punkt/.style={
           rectangle,
           rounded corners,
           draw=black, very thick,
           text width=6.5em,
           minimum height=2em,
           text centered},
    pil/.style={
           ->,
           thick,
           shorten <=2pt,
           shorten >=2pt,}
    pir/.style={
           <-,
           thick,
           shorten <=2pt,
           shorten >=2pt,}
}
\definecolor{KeynoteRed}{rgb}{.678,.051, .051}
\definecolor{KeynoteBlue}{rgb}{0.008, 0.443, 0.60}
\definecolor{KeynoteLightblue}{rgb}{.635, .914, .973}
\definecolor{KeynoteYellow}{rgb}{0.859, 0.584, 0.212}
\definecolor{KeynoteYellow}{rgb}{0.859, 0.584, 0.212}
\definecolor{KeynoteSlate}{rgb}{0.239, 0.271, 0.322}
\definecolor{KeynoteGray}{rgb}{0.498, 0.529, 0.529}
\definecolor{KeynoteGreen}{rgb}{0.18, 0.5, 0.08}
\definecolor{KeynoteTextGray}{rgb}{0.325, 0.325, 0.325}
\definecolor{KeynoteLightGray}{rgb}{0.706, 0.706, 0.706}
\definecolor{KeynoteBlueGray}{rgb}{0.471, 0.533, 0.620}
\definecolor{ECEpurple}{rgb}{.169, .18, .455}
\definecolor{ECEcyan}{rgb}{.41, .62, .72}
\definecolor{ECEgray}{rgb}{.788, .827, .859}
\definecolor{ECEblueGray}{rgb}{61.2, 70.6, 70.6}
\definecolor{ECEblueGray}{rgb}{61.2, 70.6, 70.6}
\definecolor{RiceBlue}{rgb}{0, .14, .41}

\newtoggle{ShowComments}
\togglefalse{ShowComments}

\iftoggle{ShowComments}{
	\newcommand{\evanNote}[1]{{\color{blue} {\emph{#1}}}}
	\newcommand{\evanFootnote}[1]{\footnote{{\color{blue} {\emph{#1}}}}}
	\newcommand{\johnNote}[1]{{\color{orange} {\emph{#1}}}}
	\newcommand{\jingwenNote}[1]{{\color{violet} {\emph{#1}}}}
	\newcommand{\ashuNote}[1]{{\color{red} {\emph{#1}}}}
}{
	\newcommand{\evanNote}[1]{}
	\newcommand{\evanFootnote}[1]{}
	\newcommand{\johnNote}[1]{}
	\newcommand{\jingwenNote}[1]{}
	\newcommand{\ashuNote}[1]{}
}

\newcommand{\vect}[1]{\boldsymbol{#1}}

\newcommand{\responseTx}{{A}_{T_j}}

\newcommand{\responseRx}{{A}_{R_i}}
\newcommand{\responseOneTx}{{A}_{T_1}}
\newcommand{\responseTwoTx}{{A}_{T_2}}
\newcommand{\responseOneRx}{{A}_{R_1}}
\newcommand{\responseTwoRx}{{A}_{R_2}}

\newcommand{\LenOneTx}{{L}_{T_1}}
\newcommand{\LenTwoTx}{{L}_{T_2}}
\newcommand{\LenTx}{{L}_{T_j}}
\newcommand{\LenOneRx}{{L}_{R_1}}
\newcommand{\LenTwoRx}{{L}_{R_2}}
\newcommand{\LenBS}{{L}_\mathsf{BS}}
\newcommand{\LenU}{{L}_\mathsf{Usr}}
\newcommand{\LenRx}{{L}_{R_i}}

\newcommand{\LenTwoTxPrime}{{L'}_{T_2}}

\newcommand{\LenOneRxPrime}{{L'}_{R_1}}
\newcommand{\LenOneRxPrimePrime}{{L''}_{R_1}}
\newcommand{\LenTwoRxPrime}{{L'}_{R_2}}

\newcommand{\ClusterR}[1]{\Omega_{R_{#1}}}

\newcommand{\ElevTxK}{\Theta_{T_{ij}}^{(k)}}
\newcommand{\ElevRxK}{\Theta_{R_{ij}}^{(k)}}
\newcommand{\ElevTx}{\Theta_{T_{ij}}}
\newcommand{\ElevRx}{\Theta_{R_{ij}}}
\newcommand{\ElevT}[1]{\Theta_{T_{#1}}}
\newcommand{\ElevR}[1]{\Theta_{R_{#1}}}

\newcommand{\IntT}[1]{\Psi_{T_{#1}}}
\newcommand{\IntR}[1]{\Psi_{R_{#1}}}
\newcommand{\IntTPrime}[1]{\Psi'_{T_{#1}}}
\newcommand{\IntRPrime}[1]{\Psi'_{R_{#1}}}
\newcommand{\GenieIntT}{\Psi'_{T_{2}}}
\newcommand{\GenieIntR}{\Psi'_{R_{1}}}
\newcommand{\GenieElevT}{\Theta'_{T_{2}}}
\newcommand{\GenieElevR}{\Theta'_{R_{1}}}
\newcommand{\Int}{\Psi}
\newcommand{\IntTx}{\Psi_{T_{ij}}}
\newcommand{\IntRx}{\Psi_{R_{ij}}}
\newcommand{\IntFwd}{\Psi_\mathsf{Fwd}}
\newcommand{\IntBack}{\Psi_\mathsf{Back}}
\newcommand{\TxSpace}[1]{\mathcal{T}_{#1}}
\newcommand{\TxSpacePrime}[1]{\mathcal{T}'_{#1}}
\newcommand{\RxSpace}[1]{\mathcal{R}_{#1}}
\newcommand{\RxSpacePrime}[1]{\mathcal{R}'_{#1}}
\newcommand{\RxSpacePrimePrime}[1]{\mathcal{R}''_{#1}}
\newcommand{\TxSubspace}[2]{\mathcal{T}_{#1\setminus#2}}
\newcommand{\RxSubspace}[2]{\mathcal{R}_{#1\setminus#2}}
\newcommand{\ScatOp}[1]{\mathsf{H}_{#1}}
\newcommand{\Op}[1]{\mathsf{#1}}
\newcommand{\ScatOpRes}[1]{\mathsf{H}''_{#1}}
\newcommand{\Scat}[1]{H_{#1}}
\newcommand{\TxTwoPreSpace}{\mathcal{P}_{12\setminus11}}
\newcommand{\TxIntSpace}[2]{\mathcal{T}_{#1\cap#2}}
\newcommand{\RxIntSpace}[2]{\mathcal{R}_{#1\cap#2}}
\newcommand{\RxSym}[2]{\xi_{#1}^{(#2)}}
\newcommand{\TxSym}[2]{\chi_{#1}^{\left(#2\right)}}
\newcommand{\RxVect}[1]{\boldsymbol{\xi}_{#1}}
\newcommand{\TxVect}[1]{\boldsymbol{\chi}_{#1}}
\newcommand{\RxNoise}[2]{\zeta_{#1}^{(#2)}}
\newcommand{\NoiseVect}[1]{\boldsymbol{\zeta}_{#1}}
\newcommand{\TxBasisPre}[1]{P_{12}^{(#1)}}
\newcommand{\TxBasisOrth}[1]{Q_{22\setminus12}^{(#1)}}

\newcommand{\RxBasis}[2]{J_{#1}^{(#2)}}
\newcommand{\genBasisTwo}{B_2^{(i)}}
\newcommand{\RxOneCoeffs}[1]{a_1^{(#1)}}
\newcommand{\RxTwoCoeffs}[1]{a_2^{(#1)}}
\newcommand{\Mat}[1]{\boldsymbol{A}_{#1}}
\newcommand{\SingLeft}[2]{U_{#1}^{(#2)}}
\newcommand{\SingRight}[2]{V_{#1}^{(#2)}}
\newcommand{\SingVal}[2]{\sigma_{#1}^{(#2)}}
\newcommand{\FlowOne}{\mathsf{Flow}_1}
\newcommand{\FlowTwo}{\mathsf{Flow}_2}
\newcommand{\AngleR}[1]{\theta_{R_{#1}}}
\newcommand{\AngleT}[1]{\theta_{T_{#1}}}

\newcommand{\preim}[2]{{#2}^{\leftarrow}(#1)}

\newcommand{\dPre}{d_{P}}

\newcommand{\dNull}{d_N}
\newcommand{\dInt}{d_{R\cap\mathcal{S}}}
\newcommand{\dB}{d_\mathcal{B}}
\newcommand{\rank}{\mathop{\mathrm{rank}}}
\newcommand{\imaj}{\mathrm{i}}
\newcommand{\dTwoOrth}{d_{2_\mathsf{Orth}}'}
\newcommand{\dTwoInt}{d_{2_\mathsf{Int}}'}
\newcommand{\XTwoOrth}{X_{2_\mathsf{Orth}}}
\newcommand{\XTwoInt}{X_{2_\mathsf{Int}}}

\DeclareMathOperator{\spanof}{span}

\newcommand{\DoF}{d_1 + d_2}
\newcommand{\dOneMax}{d_1^{\sf max}}
\newcommand{\dTwoMax}{d_2^{\sf max}}
\newcommand{\dSumMax}{d_{\sf sum}^{\sf max}}

\newcommand{\D}{\mathcal{D}_{\sf FD}}
\newcommand{\DHD}{\mathcal{D}_{\sf HD}}
\newcommand{\dirK}{\vect{\hat{k}}}
\newcommand{\dirKappa}{\vect{\hat{\kappa}}}

\newcounter{MYtempeqncnt}

\newtheorem{rem}{Remark}

\newtheorem{thm}{Theorem}
\newtheorem{cor}{Corollary}
\newtheorem{lem}{Lemma}
\newtheorem{define}{Definition}

\newtoggle{ZChanModel}

\begin{document}

\title{Spatial Self-Interference Isolation for In-Band Full-Duplex Wireless:\\ A Degrees-of-Freedom Analysis}

\author{Evan~Everett and Ashutosh~Sabharwal 
\thanks{E. Everett and Ashu Sabharwal are with the Department of Electrical and Computer Engineering, Rice University, Houston, TX  77005, USA. Email: evan.everett@rice.edu, ashu@rice.edu.}
\thanks{This paper was presented in part at the 2014 IEEE International Symposium on Information Theory.}
\thanks{This work was partially supported by National Science Foundation (NSF) Grants CNS 0923479, CNS 1012921, CNS 1161596 and NSF Graduate Research Fellowship 0940902.}
}

\toggletrue{ZChanModel}

\maketitle

%
\begin{abstract}
The challenge to in-band full-duplex wireless communication is managing self-interference. Many designs have employed spatial isolation mechanisms, such as shielding or multi-antenna beamforming, to isolate the self-interference wave from the receiver. Such spatial isolation methods are effective, but by confining the transmit and receive signals to a subset of the available space, the full spatial resources of the channel be under-utilized, expending a cost that may nullify the net benefit of operating in full-duplex mode. In this paper we leverage an antenna-theory-based channel model to analyze the spatial degrees~of~freedom available to a full-duplex capable base station, and observe that whether or not spatial isolation out-performs time-division (i.e. half-duplex) depends heavily on the geometric distribution of scatterers. Unless the angular spread of the objects that scatter to the intended users is overlapped by the spread of objects that backscatter to the base station, then spatial isolation outperforms time division, otherwise time division may be optimal. 
\end{abstract}


\evanNote{Comment from Evan.} \ashuNote{Comment from Ashu.} \jingwenNote{Comment from Jingwen.} \johnNote{Comment from John.}

\section{Introduction}
\label{sec:intro}

\begin{figure}[htbp]
\begin{center}
	\input{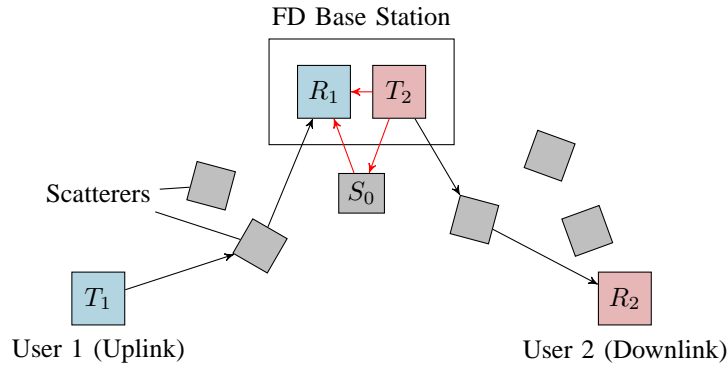}
\caption{Three-node full-duplex model}
\label{fig:threeNode}
\end{center}
\end{figure}

Currently deployed wireless communications equipment operates in half-duplex mode, meaning that transmission and reception are orthogonalized either in time (time-division-duplex) or frequency (frequency-division-duplex). Research in recent years \cite{Bliss07SimultTX_RX, Khandani2010FDPatent, Radunovic:2009aa, Duarte10FullDuplex, Choi10FullDuplex, Jain2011RealTimeFD, Duarte11FullDuplex, Sahai11FullDuplex,  Khojastepour11AntennaCancellation, Aryafar12MIDU, Duarte2012FullDuplexWiFi, Duarte12Thesis} has investigated the possibility of wireless equipment operating in full-duplex mode, meaning that transceiver will both transmit and receive at the same time and in the same spectrum.
The benefit of full-duplex is easy to see. Consider the communication scenario depicted in\johnNote{ Fig. instead of Figure throughout the manuscript.} Figure~\ref{fig:threeNode}. 
User 1 wishes to transmit uplink data to a base station, and User 2 wishes to receive downlink data from the same base station. 
If the base station is half-duplex, then it must either service the users in orthogonal time slots or in orthogonal frequency bands. 
If the base station can operate in full-duplex mode, then it can enhance spectral efficiency by servicing both users simultaneously.\footnote{We assume that the pair of users are schedules for concurrent uplink and downlink on the basis of being hidden from one another, so that interference from User~1 to User~2 will not be an issue.}
The challenge to full-duplex communication, however, is that the base station transmitter generates high-powered self-interference which potentially swamps its own receiver, precluding the reception of the uplink message. 

For full-duplex to be feasible, the self-interference must be suppressed. The two main approaches to self-interference suppression are \emph{cancellation} and \emph{spatial isolation}, and we now define each. 
Self-interference cancellation is any technique which exploits the foreknowledge of the transmit signal by subtracting an estimate of the self-interference from the received signal. The cancellation can be applied at digital baseband, at analog baseband, at RF, or, as is most common, applied at a combination of these three domains \cite{Duarte10FullDuplex, Choi10FullDuplex, Jain2011RealTimeFD, Duarte11FullDuplex, Duarte2012FullDuplexWiFi}.
Spatial isolation is any technique to spatially orthogonalize the self-interference and the signal-of-interest. Some spatial isolation techniques studied in the literature are multi-antenna beamforming \cite{Bliss07SimultTX_RX, Day12FDMIMO, Day12FDRelay}, directional antennas \cite{Everett11Asilomar}, shielding via absorptive materials \cite{Everett2013PassiveSuppressionFD}, and cross-polarization of transmit and receive antennas \cite{Everett2013PassiveSuppressionFD, Aryafar12MIDU}.  The key differentiator between cancellation and spatial isolation is that cancellation requires and exploits knowledge of the self-interference, while spatial isolation does not.
To our knowledge all full-duplex designs to date have required both cancellation and spatial isolation in order for full-duplex to be feasible even at very short ranges (i.e. $<10$ m).\footnote{For example, see designs such as \cite{Choi10FullDuplex, Duarte2012FullDuplexWiFi, Jain2011RealTimeFD, Aryafar12MIDU}, each of which leverages cancellation techniques as well as at least one spatial isolation technique.} 
Moreover, because cancellation performance is limited by transceiver impairments such as phase noise \cite{Sahai13PhaseNoise}, spatial isolation often accounts for an outsized portion of the overall self-interference suppression. 
For example, in the full-duplex design of \cite{Everett2013PassiveSuppressionFD} which demonstrated full-duplex feasibility at WiFi ranges, of the 95~dB of self-interference suppression achieved, $70$~dB is due to spatial isolation, while only $25$~dB is due to cancellation. 
Therefore if full-duplex feasibility is to be extended from WiFi-typical ranges to the ranges typical of femptocells or even larger cells, then excellent spatial isolation performance will be required, hence our focus is on spatial isolation in this paper.

In our previous work \cite{Everett2013PassiveSuppressionFD}, we studied three passive techniques for spatial isolation: directional antennas, absorptive shielding, and cross-polarization, and measured their performance in a prototype base station both in an anechoic chamber that mimics free space, and in a reflective room. As expected, the techniques suppressed the self-interference quite well (more than 70~dB) in the anechoic chamber, but in the reflective room the suppression was much less, (no more than 45 dB), due the fact that the passive techniques such as directional antennas, absorptive shielding, and cross-polarization operate primarily on the direct  path between the transmit and receive antennas, and do little to suppress paths that include an external scatterer. The direct-path limitation of passive spatial isolation mechanisms raises the question of whether or not spatial isolation can be useful in a backscattering environment. 
Another class of spatial isolation techniques called ``active'' or ``channel aware'' spatial isolation \cite{FullDuplexTutorial2014} can indeed suppress both direct an backscattered self-interference. 
In particular, if multiple antennas are used and if the self-interference channel response can be estimated, then the antenna patterns can be shaped adaptively to mitigate both direct-path and backscattered self-interference, but this pattern shaping may consume spatial resources that could have otherwise been leveraged for spatial multiplexing. Thus, there is a potential tradeoff in spatial self-interference isolation and achievable degrees of freedom.

To appreciate the tradeoff, consider the example illustrated in Figure~\ref{fig:threeNode}. 
The direct path from the base station transmitter, $T_2$, to its receiver $R_1$, can be passively suppressed by shielding the receiver from the transmitter as shown in \cite{Everett2013PassiveSuppressionFD}, but there will also be self-interference due to transmit signal backscattered from objects near the base station (depicted by gray blocks in Figure~\ref{fig:threeNode}).  The self-interference caused by scatterer $S_0$, for example, in Figure~\ref{fig:threeNode} could be avoided by creating a null in the direction of $S_0$. However losing access to that scatterer could lead to a less rich scattering environment, diminishing the spatial degrees of freedom of the uplink or downlink. 
Moreover, creating the null consumes antenna resources at the base station that could have been leveraged for spatial multiplexing to the downlink user, diminishing the spatial degrees of freedom the downlink. This example leads us to pose the following question. 
\ashuNote{Evan, maybe we need two classes of spatial isolation - active and passive. You describe passive before and state that it only handles direct path and hence ``channel-unaware" (like the terminology in our tutorial paper). The following question is interesting only if we have  channel awareness.} \evanNote{Done. See above.}

\textbf{Question}: 
Under what scattering conditions can spatial isolation be leveraged in full-duplex to provide a degree-of-freedom gain over half-duplex?
More specifically, given a constraint on the \emph{size} of the antenna arrays at the base station and at the User 1 and User 2 devices, and given a characterization of the \emph{spatial distribution} of the scatterers in the environment, what is the uplink/downlink degree-of-freedom region when the only self-interference mitigation strategy is spatial isolation?

\textbf{Modeling Approach:}
To answer the above question we leverage the antenna-theory-based channel model developed by Poon, Broderson, and Tse in \cite{TsePoon05DOF_EM,TsePoon06EmagInfoTheory,TsePoon11DOFPolarization}, which we will label the ``PBT'' model.  In the PBT model, instead of constraining the \emph{number} of antennas, the \emph{size} of the array is constrained. Furthermore,  instead of considering a channel matrix drawn from a probability distribution, a channel transfer function which depends on the geometric position of the scatterers relative to the arrays is considered. 

\textbf{Contribution:}
We extend the PBT model to the three-node full-duplex topology of Figure~\ref{fig:threeNode}, and derive the degree-of-freedom region $\D$, i.e. the set of all achievable uplink/downlink degree-of-freedom tuples. By comparing $\D$ to $\DHD$, the degree-of-freedom region achieved by time-division half-duplex, we observe that  full-duplex outperforms half-duplex, i.e. $\DHD\subset\D$, in the following two scenarios.
\begin{enumerate}
\item When the base station arrays are larger than the corresponding user arrays, 
the base station has a larger signal space than is needed for spatial multiplexing and can leverage the extra signal dimensions to form beams that avoid self-interference (i.e. self zero-forcing).
%
\ashuNote{$\leftarrow$ what do we mean by ``antennas resources'' for continuous arrays?} \evanNote{I mean a larger dimension signal space. See re-write above.}
\item More interestingly, when the forward scattering intervals and the backscattering intervals are not completely overlapped, the base station can avoid self-interference by signaling in the directions that scatter to the intended receiver, but do not backscatter to the base-station receiver. Moreover the base station can also signal in directions that \emph{do} cause self-interference, but ensure that the generated self-interference is incident on the base-station receiver only in directions in which uplink signal is \emph{not} incident on the base-station receiver,  i.e. signal such that the self-interference and uplink signal are spatially orthogonal. 

%
\end{enumerate}




\textbf{Organization of the Paper}:
Section~\ref{sec:systemModel} specifies the system model: we begin with an overview of the PBT model in Section~\ref{sec:overview} and then in Section~\ref{sec:extend} apply the model to the scenario of a full-duplex base station with uplink and downlink flows. Section~\ref{sec:analysis} gives the main analysis of the paper, the derivation of the degrees-of-freedom region. We start Section~\ref{sec:analysis} by stating the theorem which characterizes the degrees~of~freedom region and then give the achievability and converse arguments in Sections~\ref{sec:achieve}~and~\ref{sec:converse}, respectively. In Section~\ref{sec:impact} we assess the impact of the degrees-of-freedom result on the design and deployment of full-duplex base stations, and we conclude in Section~\ref{sec:conclusion}. 


\section{System Model}
\label{sec:systemModel}


We now give a brief overview of the PBT channel model presented in \cite{TsePoon05DOF_EM}. We then extend the PBT model to the case of the three-node full-duplex topology of Figure~\ref{fig:threeNode}, and define the required mathematical formalism that will ease the degrees-of-freedom analysis in the sequel. 
\subsection{Overview of the PBT Model}
\label{sec:overview}
The PBT channel model considers a wireless communication link between a transmitter equipped with a unipolarized continuous linear array of length $2L_T$ and a receiver with a similar array of length $2L_R$.
The authors observe that there are two key domains: the \emph{array domain}, which describes the current distribution on the arrays, and the \emph{wavevector domain} which describes radiated and received field patterns.
Motivated by channel measurements which show that the angles of departure and the angles of arrival of the physical paths from a transmitter to a receiver tend to be concentrated within a handful of angular clusters \cite{Poon03IndoorCharacterization,Spencer2000ClusteredChannels, CramerUWBChannels, Heddergott00NLoS_Channels}, 
the authors focus on the union of the clusters of departure angles from the transmit array, denoted $\ElevT{}$, and the union of the clusters of arrival angles to the receive array, $\ElevR{}$.
%
Because a linear array aligned to the $z$-axis array can only resolve the $z$-component, the intervals of interest are $\Psi_T = \{\cos \theta: \theta\in\ElevT{} \}$ and $\Psi_R = \{\cos \theta: \theta\in\ElevR{} \}$. In  \cite{TsePoon05DOF_EM}, it is shown from the first principles of Maxwell's equations that an array of length $2L_T$ has a resolution of $1/(2L_T)$ over the interval $\Psi_T$, so that the dimension of the transmit signal space of radiated field patterns is $2L_T|\Psi_T|$.
Likewise the dimension of the receive signal space is $2L_R|\Psi_R|$, so that the degrees of freedom of the communication link is 
\begin{equation}
d_\mathrm{P2P} = \min\left\{2L_T|\Psi_T|,  2L_R|\Psi_R| \right\}.
\end{equation}

\subsection{Extension of PBT Model to Three-Node Full-Duplex}
\label{sec:extend}

\begin{figure}[htbp]
\begin{center}
\input{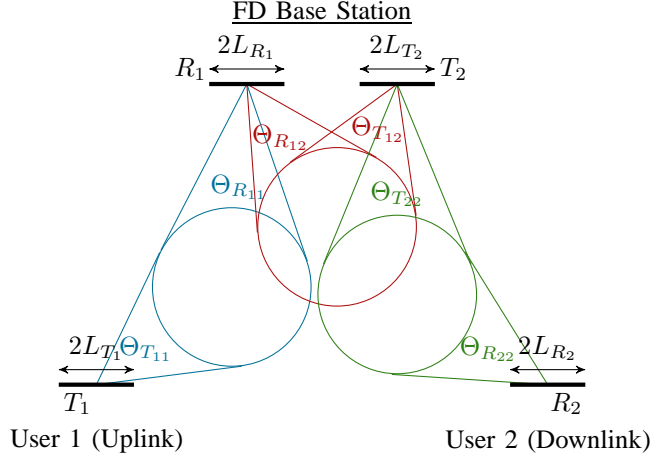}
\caption{Clustered scattering. Only one cluster for each transmit receive pair is shown to prevent clutter.
}
\label{fig:clusters}
\end{center}
\end{figure}


Now we extend the PBT channel model in \cite{TsePoon05DOF_EM}, which considers a point-to-point topology, to the three-node full-duplex topology of Figure~\ref{fig:threeNode}. Let $\FlowOne$ denote the uplink flow from User 1 to the base station, and let $T_1$ denote User~1's transmitter and $R_1$ denote the base station's receiver, as is illustrated in Figure~\ref{fig:threeNode}. Similarly, let  $\FlowTwo$  denote the downlink flow from the base station to User~2, and let $T_2$ denote the base station's transmitter and $R_2$ denote User~2's receiver.

As in \cite{TsePoon05DOF_EM}, we consider continuous linear arrays of infinitely many infinitesimally small unipolarized antenna elements. 
Each of the two transmitters $T_j,\ j = 1,2$, is equipped with a linear  array of length $2\LenTx$, and each receiver, $R_i,\ i = 1,2$, is equipped with a linear array of length $2\LenRx$. The lengths $\LenTx$ and $\LenRx$ are normalized by the wavelength of the carrier, and thus are unitless quantities. 
For each array, define a local coordinate system with origin at the midpoint of the array and $z$-axis aligned along the lengths of the array.
Let $\AngleT{j} \in [0,\pi)$ denote the elevation angle relative to the $T_j$ array, and let $\AngleR{i}$ denote the elevation angle relative to the $R_i$ array. We will see in the following that the field pattern radiated from the $T_j$ array will depend on $\AngleT{j}$ only through $\cos\AngleT{j}$. Thus let $t_j \equiv \cos\AngleT{j} \in (-1, 1]$, and likewise $\tau_i \equiv \cos\AngleR{i} \in (-1, 1]$.
Denote the current distribution on the $T_j$ array as $x_j(p_j)$, where  $p_j\in[-\LenTx,\LenTx]$ is the position along the lengths of the array, and $x_j:[-\LenTx,\LenTx] \rightarrow \mathbb{C}$ gives the magnitude and phase of the current.
The current distribution, $x_j(p_j)$, is the transmit signal controlled by $T_j$, which we constrain to be square integrable. Likewise we denote the received current distribution on the $R_i$ array as $y_i(q_i),\ q_i\in[-\LenRx,\LenRx]$. 

The current signal received by the base station receiver, $R_1$, at a point $q_1 \in [-\LenTwoRx,\LenTwoRx]$ along its array is given by
\ifCLASSOPTIONtwocolumn
\begin{align}
{y}_1(q_1) =&  
\int_{-\LenOneTx}^{\LenOneTx} {C}_{11}({q_1}, {p}_1){x}_1({p}_1)d{p}_1 
 \nonumber \\ 
&+ \int_{-\LenTwoTx}^{\LenTwoTx}{C}_{12}({q}_1,{p}_2){x}_2({p}_2)d{p}_2 
+ {z}_1({q}_1), \label{eq:y1}
\end{align}
\else
\begin{align}
{y}_1(q_1) =  
\int_{-\LenOneTx}^{\LenOneTx} {C}_{11}({q_1}, {p}_1){x}_1({p}_1)d{p}_1 
 + \int_{-\LenTwoTx}^{\LenTwoTx}{C}_{12}({q}_1,{p}_2){x}_2({p}_2)d{p}_2 
+ {z}_1({q}_1), \quad q_1 \in [-\LenOneRx,\LenOneRx] \label{eq:y1}
\end{align}
\fi
where ${z}_1({q}_1),\ q_1\in[-\LenOneRx,\LenOneRx]$ is the noise along the $R_1$ array. 
The channel response integral kernel, $C_{ij}(q_i,p_j)$, gives the current excited at a point ${q_i}$ on the $R_i$ receive array due to a current at the point $p_j$ on the $T_j$ transmit array. Note that the first term in (\ref{eq:y1}) gives the received uplink signal-of-interest, while the second term gives the self-interference generated by the base stations transmission. 
We assume that the mobile users are out of range of each other, such that there is no channel from $T_1$ to $R_2$. 
Thus $R_2$'s received signal at a point $q_2\in [-\LenTwoRx,\LenTwoRx]$ is 
\begin{align}
{y}_2({q}_2) =&  
\int_{-\LenTwoTx}^{\LenTwoTx}{C}_{22}({q}_2,{p}_2){x}_2({p}_2)d{p}_2 
+ {z}_2({q}_2), \quad q_2\in [-\LenTwoRx,\LenTwoRx]. \label{eq:y2}
\end{align} 

As in \cite{TsePoon05DOF_EM}, the channel response kernel, ${C}_{ij}(\cdot,\cdot)$, from transmitter $T_j$ to receiver $R_i$ is composed of a transmit array response $\responseTx(\cdot, \cdot)$, a scattering response $\Scat{ij}(\cdot, \cdot)$, and a receive array response $\responseRx(\cdot, \cdot)$. The channel response kernel is given by
\begin{equation}
{C}_{ij}({q},{p}) = \int \!\!\! \int \responseRx({q},\dirKappa) \Scat{ij}(\dirKappa, \dirK)  \responseTx(\dirK, {p}) d\dirK d\dirKappa,
\label{eq:channelResponse}
\end{equation}
%
where $\dirK$ is a unit vector that gives the direction of propagation from the transmitter array, and $\dirKappa$ is a unit vector that gives the direction of a propagation to the receiver array. The transmit array response kernel, $\responseTx(\dirK, {p})$, maps the current distribution along the $T_j$ array (a function of $p$) to the field pattern radiated from $T_j$ (a function of direction of departure, $\dirK$). The scattering response kernel, $\Scat{ij}(\dirKappa, \dirK)$, maps the fields radiated from $T_j$ in direction $\dirK$ to the fields incident on $R_i$ at direction $\dirKappa$. The receive array response, $\responseRx({q},\dirKappa)$, maps the field pattern incident on $R_i$ (a function of direction of arrival,  $\dirKappa$) to the current distribution excited on the $R_i$ array (a function of position $q$), which is the received signal.

\subsection{Array Responses}
In \cite{TsePoon05DOF_EM}, the transmit array response for a linear array is derived from the first principles of Maxwell's equations and shown to be 
\begin{align}
\responseTx(\dirK, {p}) = \responseTx(\cos\AngleT{j}, p) = e^{-\imaj2\pi p \cos \AngleT{j}}, \nonumber 
p \in \left[-\LenTx, \LenTx \right],
\end{align}
where $\AngleT{j}\in[0,\pi)$ is the elevation angle relative to the $T_j$ array. 
Due to the symmetry of the array (aligned to the $z$-axis) its radiation pattern is symmetric with respect to the azimuth angle and only depends on the elevation angle $\AngleT{j}$ through $\cos\AngleT{j}$. For notational convenience let $t \equiv \cos\AngleT{j} \in [-1, 1]$, so that we can simplify the transmit array response kernel to
\begin{align}
\responseTx(t, p) &= e^{-\imaj2\pi p t},\  t \in  [-1, 1],\ p \in \left[-\LenTx, \LenTx \right].
\end{align}
By reciprocity, the receive array response kernel, $\responseRx({q}, \dirKappa)$, is
\begin{align}
\responseRx(q, \tau) &= e^{\imaj2\pi q \tau},\  \tau \in  [-1, 1],\ q \in \left[-\LenRx, \LenRx \right],
\end{align}
where $\tau \equiv \cos\AngleR{i} \in [-1, 1]$ is the cosine of the elevation angle relative to the $R_i$ array. 
Note that the transmit and receive array response kernels are identical to the kernels of the Fourier transform and inverse Fourier transform, respectively, a relationship we will further explore in Section~\ref{sec:hilbertSpace}.

\subsection{Scattering Responses}
The scattering response kernel, $\Scat{ij}(\dirKappa,\dirK)$, gives the amplitude and phase of the path departing from $T_j$ in direction $\dirK$ and arriving at $R_i$ in direction $\dirKappa$. Since we are considering linear arrays which only resolve the cosine of the elevation angle, we can consider $\Scat{ij}(\tau,t)$ which gives the superposition of the amplitude and phase of all paths emanating from $T_j$ with an elevation angle whose cosine is $t$ and arriving at $R_i$ at an elevation angle whose cosine is $\tau$. 
%
%
%
As is done in \cite{TsePoon05DOF_EM}, motivated by measurements showing that scattering paths are clustered with respect to the transmitter and receiver, we adopt a model that focuses on the \emph{boundary} of the scattering clusters rather than the discrete paths themselves, as illustrated in Figure~\ref{fig:clusters}.

Let $\ElevTxK$ denote the angle subtended at transmitter $T_j$ by the $k^{\rm th}$ cluster that scatters to $R_i$, and let $\ElevTx = \bigcup_{k}\ElevTxK$ be the total transmit scattering interval from $T_j$ to $R_i$.  The scattering interval $\ElevTx$ can be thought of as the set of directions that when illuminated by $T_j$ scatters energy to $R_i$. In  Figure~\ref{fig:clusters}, to avoid clutter we illustrate the case in which $\ElevTxK$ is a single contiguous angular interval, but in general the interval will be non-contiguous and consist of several individual clusters.
Similarly let  $\ElevRxK$ denote the corresponding solid angle subtended at $R_i$ by the $k^{\rm th}$ cluster illuminated by $T_j$, and let $\ElevRx = \bigcup_{k}\ElevRxK$ be set of directions from which energy is incident on $R_i$ from $T_j$.

Thus, we see in Figure~\ref{fig:clusters} that from the point-of-view of the base-station transmitter, $T_2$, $\ElevT{22}$ is the angular interval over which it can radiate signals that will couple to the intended downlink receiver, $R_2$,  while $\ElevT{12}$ is the interval in which radiated signals will scatter back 
to the base station receiver, $R_1$, as self-interference. Likewise, from the point-of-view of the base station receiver, $R_1$,  $\ElevR{11}$ is the interval over which it may receive signals from the User~1 transmitter, $T_1$, while $\ElevR{12}$ is the interval in which self-interference may be present. Clearly, the extent to which the interference intervals and the signal-of-interest intervals overlap will have a major impact on the degrees of freedom of the network. 
Because linear arrays can only resolve the cosine of the elevation angle $t\equiv\cos\theta$, let us denote the ``effective'' scattering interval as 
$$\IntTx \equiv \left\{t: \arccos(t) \in \ElevTx  \right\} \subset [-1,1].$$
Likewise for the receiver side we denote the effective scattering intervals as
$$\IntRx \equiv \left\{\tau: \arccos(\tau) \in \ElevRx  \right\} \subset [-1,1].$$ Define the size of the transmit and receive scattering intervals as
\begin{equation}
|\IntTx| = \int_{\IntTx} t\, dt,\qquad |\IntRx| = \int_{\IntRx} \tau\, d\tau.
\end{equation}
As in \cite{TsePoon05DOF_EM}, we assume the following characteristics of the scattering responses: 
\begin{enumerate}

\item $\Scat{ij}(\tau,t) \neq 0$ only if $(\tau,t) \in \IntRx\times \IntTx$.
 

\item $\int||\Scat{ij}(\tau,t)||dt \neq 0\ \forall \ \tau \in \IntRx$.

\item $\int||\Scat{ij}(\tau,t)||d\tau \neq 0\ \forall\ t \in \IntTx$.

\item The point spectrum of $\Scat{ij}(\cdot,\cdot)$, excluding $0$, is infinite.

\item $\Scat{ij}(\cdot,\cdot)$ is Lebesgue measurable, that is 
$\int_{-1}^1 \int_{-1}^1 |\Scat{ij}(\tau,t)|^2 \,d\tau \,dt < \infty.$


\end{enumerate}
The first condition means that the scattering response is zero unless the angle of arrival and angle of departure both lie within their respective  scattering intervals.  
The second condition means that in any direction of departure, $t \in \IntTx$, from $T_j$ there exists at least one path to receiver $R_i$. Similarly, the third condition implies that in any direction of arrival, $\tau \in \IntRx$, to $R_i$ there exists at least one path from $T_j$. 
The fourth condition means that there are many paths from the transmitter to the receiver within the scattering intervals, so that the number of propagation paths that can be resolved within the scattering intervals is limited by the length of the arrays and not by the number of paths. The final condition aids our analysis by ensuring the corresponding integral operator is compact, but is also physically justified assumption since one could argue for the stricter assumption $\int_{-1}^1 \int_{-1}^1 |\Scat{ij}(\tau,t)|^2 \,d\tau \,dt \leq 1$, since no more energy can be scattered than is transmitted. 


\subsection{Hilbert Space of Wave-vectors}
\label{sec:hilbertSpace}
We can now write the original input-output relation given in (\ref{eq:y1})~and~(\ref{eq:y2}) as
\begin{align}
y_1(q) 	 &= 	\int_{\IntR{11}} \!\!\!\!\!\!\!\! \responseOneRx(q,\tau) 
			\int_{\IntT{11}} \!\!\!\!\!\!\!\! H_{11}(\tau,t) 
			\int_{-\LenOneTx}^{\LenOneTx}\!\!\!\!\!\!\!\! 
			\responseOneTx(t,p) x_1(p)\,  d\tau\, dt\, dp \nonumber \\
	   &+ 	\int_{\IntR{12}} \!\!\!\!\!\!\!\! \responseOneRx(q,\tau) 
			\int_{\IntT{12}} \!\!\!\!\!\!\!\! H_{12}(\tau,t) 
			\int_{-\LenTwoTx}^{\LenTwoTx}\!\!\!\!\!\!\!\! 
			\responseTwoTx(t,p) x_2(p)\,  d\tau\, dt\, dp + z_1(q), \label{eq:Zchan1}\\
y_2(q) 	&= 	\int_{\IntR{22}} \!\!\!\!\!\!\!\! \responseTwoRx(q,\tau) 
			\int_{\IntT{22}} \!\!\!\!\!\!\!\! H_{22}(\tau,t) 
			\int_{-\LenTwoTx}^{\LenTwoTx}\!\!\!\!\!\!\!\! 
			\responseTwoTx(t,p) x_2(p)\,  d\tau\, dt\, dp + z_2(q) \label{eq:Zchan2}.
\end{align}
The channel model of  (\ref{eq:Zchan1})~and~(\ref{eq:Zchan2}) is expressed in the \emph{array} domain, that is the transmit and receive signals are expressed as the current distributions excited along the array. Just as one can simplify a signal processing problem by leveraging the Fourier integral to transform from the time domain to the frequency domain, we can leverage the transmit and receive array responses to transform the problem from the array domain to the \emph{wave-vector} domain. In other words, we can express the transmit and receive signals as field distributions over direction rather than current distributions over position along the array. In fact, for our case of the unipolarized linear array, the transmit and receive array responses \emph{are} the Fourier and inverse-Fourier integral kernels, respectively. 

Let $\TxSpace{j}$ be the space of all field distributions that transmitter $T_j$'s array of length $\LenTx$ can radiate towards the available scattering clusters, $\IntT{jj}\cup\IntT{ij}$ (both signal-of-interest and self-interference). In the vernacular of \cite{TsePoon05DOF_EM}, $\TxSpace{j}$ is the space of field distributions array-limited to $\LenTx$ and wavevector-limited to $\IntT{jj}\cup\IntT{ij}$.
To be precise, define $\TxSpace{j}$ to be the Hilbert space of all square-integrable functions  $X_j:\IntT{jj}\cup\IntT{ij}\rightarrow \mathbb{C}$, that can be expressed as
$$
X_j(t) = \int_{-\LenTx}^{\LenTx} \responseTx(t,p) x_j(p)\, dp,\ \quad t\in \IntT{jj}\cup\IntT{ij}
$$
for some $x_j(p),\ p\in [-\LenTx,\LenTx]$. The inner product between two member functions, $U_j,V_j\in\TxSpace{j}$, is the usual inner product
$$
\langle U_j,V_j\rangle = \int_{\IntT{jj}\cup\IntT{ij}} U_j(t)V_j^*(t)\, dt.
$$
Likewise let $\RxSpace{i}$ be the space of field distributions that can be incident on receiver $R_i$ from the available scattering clusters, $\IntR{ii}\cup\IntR{ij}$, and resolved by an array of length $\LenRx$.
More precisely, $\RxSpace{i}$ is the Hilbert space of all square-integrable functions  $Y_i:\IntR{ii}\cup\IntR{ij} \rightarrow \mathbb{C}$, that can be expressed as  
$$
Y_i(\tau) = \int_{-\LenRx}^{\LenRx} \responseRx^*(q,\tau) y_i(q)\, dq,\ \quad \tau\in \IntR{ii}\cup\IntR{ij}
$$
for some $y_i(q),\ q\in [-\LenRx,\LenRx]$, with the usual inner product. 
From \cite{TsePoon05DOF_EM}, we know that the dimension of these array-limited and wavevector-limited transmit and receive spaces are, respectively, 
\begin{align}
\dim \TxSpace{j} &= 2\LenTx |\IntT{jj}\cup\IntT{ij}|\text{, and} \\
\dim \RxSpace{i} &= 2\LenRx |\IntR{ii}\cup\IntR{ij}|.
\end{align}
We can think of the scattering integrals in (\ref{eq:Zchan1})~and~(\ref{eq:Zchan2}) as operators mapping from one Hilbert space to another. Define the operator $\ScatOp{ij}:\TxSpace{j}\rightarrow\RxSpace{i}$ by
\begin{equation}
(\ScatOp{ij}X_j)(\tau) = \int_{\IntT{ij} \cup \IntT{jj}} \!\!\!\!\!\!\!\!\! \Scat{ij}(\tau,t) X_j(t)\,dt,\  \tau \in \IntR{ij}\cup \IntR{ii}.
\end{equation}
We can now write the channel model of (\ref{eq:Zchan1})~and~(\ref{eq:Zchan1}) in the wave-vector domain as
\begin{align}
Y_1 &= \ScatOp{11}X_1 + \ScatOp{12}X_2 + Z_2, \label{eq:Y1}\\
Y_2 &= \ScatOp{22}X_2 + Z_2,\label{eq:Y2}
\end{align}
where $X_j \in \RxSpace{j}$, for $j=1,2$ and $Y_i, Z_i \in \RxSpace{i}$ for $i=1,2$. 

The following lemma states key properties of the scattering operators in~(\ref{eq:Y1}-\ref{eq:Y2}), that we will leverage in our analysis. 

\begin{lem}
\label{lem:scatteringProperties}
The scattering operators $\ScatOp{ij}$, $(i,j)\in\{(1,1), (2,2), (1,2)\}$ have the following properties:
\begin{enumerate}
\item $\ScatOp{ij}: \TxSpace{j}\rightarrow\RxSpace{i}$ is a compact operator
\item 
	$
	\dim R(\ScatOp{ij}) = \dim N(\ScatOp{ij})^\perp \nonumber =  2\min\{\LenTx |\IntTx|, \LenRx |\IntRx| \}
	$
\item There exists a singular system $\left\{\SingVal{ij}{k}, \SingLeft{ij}{k}, \SingRight{ij}{k}\right\}_{k=1}^{\infty}$ for operator  $\ScatOp{ij}$, and a singular value $\SingVal{ij}{k}$ is nonzero if and only if $k\leq 2\min\{\LenTx |\IntTx|, \LenRx |\IntRx| \}$. 
\end{enumerate}
\end{lem}
\begin{IEEEproof}
Property 1 holds because we have assumed that $\Scat{ij}(\cdot,\cdot)$, the  kernel of integral operator $\ScatOp{ij}$, is square integrable, and any  integral operator with a square integrable kernel is compact (see Theorem~8.8 of \cite{Young88Hilbert}). Property 2 is just a restatement of the main result of \cite{TsePoon05DOF_EM}. Property 3 follows from the first two properties: The compactness of $\ScatOp{ij}$, established in Property 1, implies the existence of a singular system, since there exists a singular system for any compact operator (see Section~16.1 of \cite{Young88Hilbert}). Property 2 implies that only the first $2\min\{\LenTx |\IntTx|, \LenRx |\IntRx| \}$ of the singular values will be nonzero, since the $\{\SingLeft{ij}{k}\}$ corresponding to nonzero singular values form a basis for $R(\ScatOp{ij})$, which has dimension $2\min\{\LenTx |\IntTx|, \LenRx |\IntRx| \}$ .  See Lemma~\ref{lem:SVD} in Appendix~\ref{subsec:lems} for a description of the properties of singular systems for compact operators, or see Section 2.2 of \cite{Engl:InverseProblemsBook} or Section 16.1 of \cite{Young88Hilbert} for a thorough treatment.
\end{IEEEproof}


\section{Spatial Degrees-of-Freedom Analysis}
\label{sec:analysis}
We now give the main result of the paper: a characterization of the spatial degrees-of-freedom region for the PBT channel model applied full-duplex base station with uplink and downlink flows.
\begin{thm}
\label{thm:mainResult}
Let $d_1$ and $d_2$ be the spatial degrees of freedom of $\FlowOne$ and $\FlowTwo$ respectively. The spatial degrees-of-freedom region, $\D$, of the three-node full-duplex channel is the convex hull of all spatial degrees-of-freedom tuples, $(d_1,d_2)$, satisfying
\begin{align}
d_1 \leq &\ \dOneMax =  2\min ( \LenOneTx |\IntT{11}|, \LenOneRx |\IntR{11}| ), \label{eq:d1Bound}\\
d_2 \leq &\ \dTwoMax = 2 \min ( \LenTwoTx |\IntT{22}|, \LenTwoRx |\IntR{22}| ), \label{eq:d2Bound}\\
\ifCLASSOPTIONtwocolumn
{d_1 + d_2} \leq  &\ \dSumMax =
2\LenTwoTx |\IntT{22} \setminus \IntT{12}| + 2\LenOneRx |\IntR{11} \setminus \IntR{12}|  
\nonumber \\ & \qquad \qquad 
+2 \max (\LenTwoTx |\IntT{12}|,  \LenOneRx |\IntR{12}|) . \label{eq:dSumBound}
\else
{d_1 + d_2} \leq  &\ \dSumMax =
2\LenTwoTx |\IntT{22} \setminus \IntT{12}| + 2\LenOneRx |\IntR{11} \setminus \IntR{12}| 
+2 \max (\LenTwoTx |\IntT{12}|,  \LenOneRx |\IntR{12}|) . \label{eq:dSumBound}
\fi
%
%
\end{align}
\end{thm}
The degrees-of-freedom region characterized by Theorem~\ref{thm:mainResult} $\D$ is the pentagon-shaped region shown in Figure~\ref{fig:region}. 
The achievability part of Theorem~\ref{thm:mainResult} is given in Section~\ref{sec:achieve} and the converse is given in Section~\ref{sec:converse}. 

\begin{figure}[htbp]
\begin{center}
	\begin{tikzpicture}
\begin{axis}[%
scale only axis,
width=2.3in, height=1.8in, 
xmin=0, xmax=6,
ymin=0, ymax=5,
xlabel={$d_1$},
ylabel={$d_2$},
ylabel near ticks,
xtick={1,2,3,4,5},
xticklabels={,{ \footnotesize $\dSumMax - \dOneMax$},,$\dOneMax$,},
ytick={1,2,3,4,5},
yticklabels={,{\footnotesize $\dSumMax - \dTwoMax$},,$\dTwoMax$,},
]
\addplot [color=blue, only marks, mark = *] coordinates{
	(2, 4)
	(4, 2)
};%
\addplot [color=black] coordinates{
	(4, 0)
	(4, 2)
	(2, 4)
	(0, 4)
};
\addplot [color=black, dashed] coordinates{
	(2, 0)
	(2, 4)
};%
\addplot [color=black, dashed] coordinates{
	(0, 2)
	(4, 2)
};%
\node[pin=60:{\small $d_1+d_2 = \dSumMax$}] at (axis description cs:0.50,0.55) {};
\node[] at (axis description cs:0.44,0.87) {\small$(d_1'', d_2'')$};
\node[] at (axis description cs:0.78,0.4) {\small$(d_1', d_2')$};
\end{axis}
\end{tikzpicture}
\caption{degrees-of-freedom region, $\D$}
\label{fig:region}
\end{center}
\end{figure}

	\subsection{Achievability}
	\label{sec:achieve}

We establish achievability of $\D$ by way of two lemmas. The first lemma shows the achievability of two specific spatial degrees-of-freedom tuples, and the second lemma shows that these tuples are indeed the corner points of $\D$. 

\begin{lem}
\label{thm:achievePoints}
The spatial degree-of-freedom tuples $(d_1',d_2')$ and $(d_1'',d_2'')$ are achievable, where 

\begin{align}
d_1' =&\min\left\{2\LenOneTx |\IntT{11}|, 2\LenOneRx |\IntR{11}|\right\}, \label{eq:point1} \\
d_2' =& \min\left\{d_{T_2},2\LenTwoRx |\IntR{22}|\right\} 1(\LenOneTx |\IntT{11}| \geq \LenOneRx |\IntR{11}|) \nonumber \\ 
 &+ \min\left\{\delta_{T_2},2\LenTwoRx |\IntR{22}|\right\} 1(\LenOneTx |\IntT{11}| < \LenOneRx |\IntR{11}|), \label{eq:point2}\\
d_1'' =& \min\left\{2\LenOneTx |\IntT{11}|, d_{R_1}\right\} 1(\LenTwoRx |\IntR{22}| \geq \LenTwoTx |\IntT{22}|) \nonumber \\ 
 &+ \min\left\{2\LenOneTx |\IntT{11}|, \delta_{R_1} \right\} 1(\LenTwoRx |\IntR{22}| < \LenTwoTx |\IntT{22}|),\\
d_2'' =&\min\left\{2\LenTwoTx |\IntT{22}|, 2\LenTwoRx |\IntR{22}|\right\}, 
\end{align}
with $d_{T_2}$, $\delta_{T_2}$, $d_{R_1}$, and $\delta_{R_1}$ given in (\ref{eq:dT2}-\ref{eq:deltaR1}), and where $1(\mathsf{arg})$ is an indicator function that evaluates to one if the argument it true, and otherwise evaluates to zero. 
\evanNote{Reminder: check before submitting. Equations number are set manually. So may be wrong is equations are added or deleted before this page.} 
%
%

\begin{figure*}[!h]
\normalsize
\setcounter{MYtempeqncnt}{\value{equation}}
\setcounter{equation}{21}
\hrulefill
\begin{align}
d_{T_2} = 2 \LenTwoTx |\IntT{22} \setminus \IntT{12}|
+ 2\min &\left\{   \LenTwoTx |\IntT{22} \cap \IntT{12}|,  
(\LenTwoTx |\IntT{12}| - \LenOneRx |\IntR{12}| )^+  +   \LenOneRx |\IntR{12} \setminus \IntR{11}| \right\} \label{eq:dT2}\\
%
%
\delta_{T_2} = 2\LenTwoTx |\IntT{22} \setminus \IntT{12}| +  2\min &\left\{ \vphantom{\left[|\IntT{12}|^+ \right]} \LenTwoTx |\IntT{22} \cap \IntT{12}|,\, \LenTwoTx |\IntT{12}|   
\right. \nonumber \\  &\  \left.
-\left[\LenOneTx|\IntT{11}| - \left(\LenOneRx |\IntR{11} \setminus \IntR{12}| + (\LenOneRx |\IntR{12}| - \LenTwoTx |\IntT{12}|)^+ \right) \right]  \right\} \label{eq:deltaT2}
%
\end{align}
\begin{align}
d_{R_1} = 2\LenOneRx |\IntR{11} \setminus \IntR{12}| + 2\min &\left\{   \LenOneRx|\IntR{11} \cap \IntR{12}|,
(\LenOneRx|\IntR{12}| - \LenTwoTx|\IntT{12}|)^+ + \LenTwoTx|\IntT{12} \setminus \IntT{22}|\right\}\label{eq:dR1}
\\
\delta_{R_1} = 2 \LenOneRx |\IntR{11} \setminus \IntR{12}| + 2\min &\left\{ \vphantom{\left[|\IntT{12}|^+ \right]}  \LenOneRx |\ClusterR{11} \cap \IntR{12}|,\ 
\LenOneRx |\IntR{12}| 
\right. \nonumber \\  &  \left.
- \left[\LenTwoRx|\IntR{22}| - \left(\LenTwoTx |\IntT{22} \setminus \IntT{12}| + (\LenTwoTx |\IntT{12}| - \LenOneRx |\IntR{12}|)^+ \right) \right] \right\} \label{eq:deltaR1}
\end{align}
\setcounter{equation}{\value{MYtempeqncnt}}
\hrulefill
\vspace*{4pt}
\end{figure*}
\end{lem}
\setcounter{equation}{25}
	\begin{IEEEproof}
Due to the symmetry of the problem, it suffices to demonstrate achievability of only the first spatial degree-of-freedom pair in Lemma~\ref{thm:achievePoints}, ($d_1', d_2'$), as the second pair, ($d_1'', d_2''$), follows from the symmetry.
Thus we seek to prove the achievability of the tuple ($d_1', d_2'$) given in (\ref{eq:point1}-\ref{eq:point2}). 
We will show achievability of ($d_1', d_2'$) in the case where $\LenOneTx |\IntT{11}| \geq \LenOneRx |\IntR{11}|$, for which
\begin{align}
d_1' &=2\LenOneRx |\IntR{11}|, \label{eq:d1PrimeCase1} \\
d_2' &=\min\left\{d_{T_2},2\LenTwoRx |\IntR{22}|\right\}, \label{eq:d2PrimeCase1}
\end{align}
where
\ifCLASSOPTIONtwocolumn
\begin{align}
&d_{T_2} =  2\LenTwoTx |\IntT{22} \setminus \IntT{12}| + \nonumber \\
& \min \left\{ \!\!\!
\begin{tabular}{c}
$2\LenTwoTx |\IntT{22} \cap \IntT{12}|$, \tabularnewline 
$2(\LenTwoTx |\IntT{12}| - \LenOneRx |\IntR{12}| )^+  +   2\LenOneRx |\IntR{12} \setminus \IntR{11}|$
\end{tabular} 
\!\!\! \right\}. \label{eq:T2DOF}
\end{align}
\else
\begin{align}
&d_{T_2} =  2\LenTwoTx |\IntT{22} \setminus \IntT{12}| +  \min \left\{ \!\!\!
\begin{tabular}{c}
$2\LenTwoTx |\IntT{22} \cap \IntT{12}|$, \tabularnewline 
$2(\LenTwoTx |\IntT{12}| - \LenOneRx |\IntR{12}| )^+  +   2\LenOneRx |\IntR{12} \setminus \IntR{11}|$
\end{tabular} 
\!\!\! \right\}. \label{eq:T2DOF}
\end{align}
\fi
Achievability of ($d_1', d_2'$) in the $\LenOneTx |\IntT{11}| < \LenOneRx |\IntR{11}|$ case is analogous.\evanFootnote{This also may not obvious, it could be good to write down that case also, but put it in the appendix.}
%
We now begin the steps to show achievability of (\ref{eq:d1PrimeCase1}-\ref{eq:d2PrimeCase1}).
	\subsubsection{Defining Key Subspaces} We first define some subspaces of the transmit and receive wave-vector spaces ($\TxSpace{1}$, $\TxSpace{2}$, $\RxSpace{1}$, and $\RxSpace{2}$) that will be crucial in demonstrating achievability.

\textbf{Subspaces of $\TxSpace{2}$}: 
Recall that $\TxSpace{2}$ is the space of all field distributions that can be radiated by the base station transmitter, $T_2$, in the direction of the scatterer intervals, $\IntT{22} \cup \IntT{12}$, (both signal-of-interest and self-interference). 
Let $\TxSubspace{22}{12} \subseteq \TxSpace{2}$ be the subspace of field distributions that can be transmitted by $T_2$, which are nonzero only in the interval  $\IntT{22}\setminus \IntT{12}$,
\begin{equation}
\TxSubspace{22}{12} \equiv  \spanof \{ X_2 \in \TxSpace{2}: X_2(t) = 0\ \forall\ t \in \IntT{12} \}.
\end{equation}
More intuitively, $\TxSubspace{22}{12}$ is the space of transmissions from the base station which couple only to the intended downlink user, and do not couple back to the base station receiver as self-interference.  
Similarly let $\TxSpace{12} \subseteq \TxSpace{2}$ the subspace of functions that are only nonzero in the interval $\IntT{12}$, 
\begin{equation}
\TxSpace{12} \equiv  \spanof \{ X_2 \in \TxSpace{2}: X_2(t) = 0\ \forall\ t \notin \IntT{12} \},
\end{equation}
that is, the space of base station transmissions which \emph{do} couple to the base station receiver as self-interference.
Finally, let $\TxIntSpace{22}{12}\subseteq \TxSpace{12} \subseteq \TxSpace{2}$ be the subspace of field distributions that are nonzero only in the interval  $\IntT{22}\cap \IntT{12}$,
\begin{equation}
\TxIntSpace{22}{12}\equiv  \spanof \{ X_2 \in \TxSpace{2}: X_2(t) = 0\ \forall\ t \notin \IntT{22} \cap \IntT{12} \},
\end{equation}
the space of base station transmission which couple \emph{both} to the downlink user and to the base station receiver.
From the result of \cite{TsePoon05DOF_EM}, we know that the dimension of each of these transmit subspaces of $\TxSpace{1}$ is as follows:
\begin{align}
\dim \TxSpace{12} &= 2\LenTwoTx |\IntT{12}|,\\
\dim \TxSubspace{22}{12} &= 2\LenTwoTx |\IntT{22} \setminus \IntT{12}|, \label{eq:dimOrth}\\
\dim\TxIntSpace{22}{12} &= 2\LenTwoTx |\IntT{22} \cap \IntT{12}|\label{eq:dimInt}. 
\end{align}
One can check that $\TxSpace{12}$ and  $\TxSubspace{22}{12}$ are constructed such that they form an \emph{orthogonal direct sum} for space $\TxSpace{2}$, a relation we notate as
\begin{equation}
 \TxSpace{2} = \TxSpace{12} \oplus \TxSubspace{22}{12}.
\end{equation}
By \emph{orthogonal direct sum} we mean that any $X_2 \in \TxSpace{2}$ can be written as $X_2 = \XTwoOrth + \XTwoInt$, for some  $\XTwoOrth\in \TxSubspace{22}{12}$ and $\XTwoInt\in \TxSpace{12}$, such that $\XTwoOrth \perp \XTwoInt.$
By the construction of $\TxSubspace{22}{12}$, $\ScatOp{12}\XTwoOrth = 0$, since $\Scat{12}(\tau,t) = 0$ $\forall\, t\notin \IntT{12}$ and $\XTwoOrth\in \TxSubspace{22}{12}$ implies $\XTwoOrth(t) = 0\ \forall\ t \in \IntT{12}$.  In other words, $\XTwoOrth\in \TxSubspace{22}{12}$ is zero everywhere the integral kernel $\Scat{12}(\tau,t)$ is nonzero. Thus any transmitted field distribution that lies in the subspace $\TxSubspace{22}{12}$ will not present any interference to $R_2$.

\textbf{Subspaces of $\TxSpace{1}$}: 
Recall that $\TxSpace{1}$ is the space of all field distributions that can be radiated by the uplink user transmitter, $T_1$, towards the available scatterers. 
Let $\TxSpace{11} \subseteq \TxSpace{1}$ be the subspace of field distributions that can be transmitted by $T_1$'s continuous linear array of length $\LenOneTx$ which are nonzero only in the interval $\IntT{11}$,\footnote{Note that $\TxSpace{11}=\TxSpace{1}$, since we have assumed $\IntT{21} = \emptyset$.  Although $\TxSpace{11}$ is thus redundant, we define it for notational consistency} more precisely
\begin{equation}
\TxSpace{11} \equiv  \spanof \{ X_1 \in \TxSpace{1}: X_1(t) = 0\ \forall\ t \notin \IntT{11} \}.
\end{equation}
More intuitively, $\TxSpace{11}$ is the space of transmissions from the uplink user which will couple to the base station receiver. From the result of \cite{TsePoon05DOF_EM}, we know that
\begin{align}
\dim \TxSpace{11} = 2\LenOneTx |\IntT{11}|\label{eq:dimTOne}. 
\end{align}

\textbf{Subspaces of $\RxSpace{1}$}: 
Recall that $\RxSpace{1}$ is the space of all incident field distributions that can be resolved by the base station receiver, $R_1$.
Let $\RxSpace{12} \subseteq \RxSpace{1}$ to be the subspace of received field distributions  which are nonzero only for $\tau \in \IntR{12}$, that is
\begin{equation}
\RxSpace{12} \equiv  \spanof \{ Y_1 \in \RxSpace{1}: Y_1(\tau) = 0\ \forall\ \tau \notin \IntR{12} \}.
\end{equation}
Less formally, $\RxSpace{12}$ is the space of receptions at the base station which could have emanated from the base stations own transmitter.  
Similarly $\RxSubspace{12}{11}\subseteq \RxSpace{12} \subseteq \RxSpace{1}$ be the subspace of received field distributions that are only nonzero for $\tau \in \IntR{12} \setminus \IntR{11}$, 
\begin{equation}
\RxSubspace{12}{11} \equiv \spanof \{ Y_1 \in \RxSpace{1}: Y_1(\tau) = 0\ \forall\ \tau \in \IntR{11} \}.
\end{equation}
Less formally,  $\RxSubspace{12}{11}$ is the space of receptions at the base station which could have emanated from the base station transmitter, but could not have emanated from the uplink user. 
Finally, define $\RxSpace{11} \subseteq \RxSpace{1}$ to be the subspace of received field distributions that are nonzero only for $\tau \in \IntR{11}$, 
\begin{equation}
\RxSpace{11} \equiv  \spanof \{ Y_1 \in \RxSpace{1}: Y_1(\tau) = 0\ \forall\ \tau \notin \IntR{11} \},
\end{equation}
the space of base station receptions which could have emanated from the intended uplink user. 
Note that 
$
\RxSpace{1} = \RxSpace{11} \oplus \RxSubspace{12}{11}.
$ 
From the result of \cite{TsePoon05DOF_EM}, we know the dimension of each of the above base-station receive subspaces is as follows:
\begin{align}
\dim \RxSpace{11} &= 2\LenOneRx |\IntR{11}|,\\
\dim \RxSubspace{12}{11} &= 2\LenOneRx |\IntR{12} \setminus \IntR{11}|,\\
\dim \RxSpace{12} &= 2\LenOneRx |\IntR{12}|.
\end{align} 

\textbf{Subspaces of $\RxSpace{2}$}: 
Recall that $\RxSpace{2}$ is the space of all incident field distributions that can be resolved by the downlink user receiver, $R_2$. 
Let $\RxSpace{22} \subseteq \RxSpace{2}$ to be the subspace of received field distributions which are nonzero only for $\tau \in \IntR{22}$,\footnote{Note that $\RxSpace{22}=\RxSpace{2}$, since we have assumed $\IntR{21} = \emptyset$.  Although $\RxSpace{22}$ is thus redundant, we define it for notational consistency} that is
\begin{equation}
\RxSpace{22} \equiv  \spanof \{ Y_2 \in \RxSpace{2}: Y_2(\tau) = 0\ \forall\ \tau \notin \IntR{22} \}.
\end{equation}

By substituting the subspace dimensions given above into  (\ref{eq:d1PrimeCase1}-\ref{eq:T2DOF}),
we can restate the degree-of-freedom pair whose achievability we are establishing as 
\begin{align}
d_1' &=\dim \RxSpace{11}, \\
d_2' &=\min\left\{d_{T_2},\dim \RxSpace{22} \right\},
\end{align}
where

\ifCLASSOPTIONtwocolumn
\begin{align}
d_{T_2} &= \dim \TxSubspace{22}{12}  \\
&+ \min \left\{ \!\!\!
\begin{tabular}{c}
$\dim \TxIntSpace{22}{12}$, \tabularnewline 
$(\dim \TxSpace{12} - \dim \RxSpace{12} )^+  + \dim \RxSubspace{12}{11}$
\end{tabular} 
\!\!\! \right\}. \nonumber
\end{align}
\else
\begin{align}
d_{T_2} &= \dim \TxSubspace{22}{12}  + \min \left\{ \!\!\!
\begin{tabular}{c} 
$\dim \TxIntSpace{22}{12}$, \tabularnewline 
$(\dim \TxSpace{12} - \dim \RxSpace{12} )^+  + \dim \RxSubspace{12}{11}$
\end{tabular} 
\!\!\! \right\}. \label{eq:dTwoAchieve}
\end{align}
\fi
Now that we have defined the relevant subspaces, we can show how these subspaces are leveraged in the transmission and reception scheme that achieves the spatial degrees-of-freedom tuple $(d_1', d_2')$. 

\subsubsection{Spatial Processing at each Transmitter/Receiver}
We now give the transmission schemes at each transmitter, and the recovery schemes at each receiver. 

\textbf{Processing at uplink user transmitter, $T_1$:}
Recall that  $d_1' = \dim \RxSpace{11}$ is the number of spatial degrees-of-freedom we wish to achieve for $\FlowOne$, the uplink flow. Let
$$\left\{ \TxSym{1}{k} \right\}_{k=1}^{d_1'},\ \TxSym{1}{i} \in \mathbb{C},$$ be the $d_1'$ symbols that $T_1$ wishes to transmit to $R_1$. We know from Lemma~\ref{lem:scatteringProperties} there exists and singular value expansion for $\ScatOp{11}$, so let  
$$\left\{\SingVal{11}{k}, \SingLeft{11}{k}, \SingRight{11}{k}\right\}_{k=1}^{\infty}$$ be a singular system\footnote{See Lemma~\ref{lem:SVD} in Appendix~\ref{subsec:lems} for the definition of a singular system.} for the operator $\ScatOp{11}: \TxSpace{1}\rightarrow \RxSpace{1}$. 
Note that the functions 
$$ \left\{ \SingRight{11}{k} \right\}_{k=1}^{\dim \TxSpace{1}}$$ form an orthonormal basis for $\TxSpace{1}$, and since $d_1' = \dim \RxSpace{11}  \leq \dim \TxSpace{1}$, there are at least as many such basis functions as there are symbols to transmit.
%
We construct $X_1$, the transmit wave-vector signal transmitted by $T_1$, as
\begin{equation}
X_1 = \sum_{k=1}^{d_1'} \TxSym{1}{k} \SingRight{11}{k}.
\end{equation}

\textbf{Processing at the base station transmitter, $T_2$:}
%
Recall that  $d_2' =\min\left\{d_{T_2},2\LenTwoRx |\IntR{22}|\right\}$, where $d_{T_2}$ is given in (\ref{eq:dTwoAchieve}), is the number of spatial degrees-of-freedom we wish to achieve for $\FlowTwo$, the downlink flow.
Let
$$\left\{ \TxSym{2}{k} \right\}_{k=1}^{d_2'}$$
be the $d_2'$ symbols that $T_2$ wishes to transmit to $R_2$.
We split the $T_2$ transmit signal into the sum of two orthogonal components, $\XTwoOrth \in \TxSubspace{22}{12} $ and $\XTwoInt \in \TxSpace{12}$, so that the wave-vector signal transmitted by $T_2$ is
\begin{equation}
X_2 = \XTwoOrth + \XTwoInt,\quad \XTwoOrth \in \TxSubspace{22}{12},\quad \XTwoInt\in\TxSpace{12}.
\end{equation}
Recall that $\XTwoOrth\in\TxSubspace{22}{12}$ implies $\ScatOp{12}\XTwoOrth = 0$. Thus we can construct $\XTwoOrth \in \TxSubspace{22}{12}$ without regard to the structure of $\ScatOp{12}$.
Let 
$$\left\{ \TxBasisOrth{i} \right\}_{i=1}^{\dim \TxSubspace{22}{12}}$$
be an arbitrary orthonormal basis for $\TxSubspace{22}{12}$,
and let 
\begin{equation}
\dTwoOrth \equiv \min\left\{\dim \TxSubspace{22}{12}, \dim \RxSpace{22} \right\},
\end{equation}
be the number of symbols that $T_2$ will transmit along $\XTwoOrth$.
We construct $\XTwoOrth$ as  
\begin{align}
\XTwoOrth = \sum_{i=1}^{\dTwoOrth}\TxSym{2}{i} \TxBasisOrth{i}.
\end{align}

Recall that there are $d_2'$ total symbols that $T_2$ wishes to transmit, and we have transmitted $\dTwoOrth$ symbols along $\XTwoOrth$, thus there are $d_2' -\dTwoOrth$ symbols remaining to transmit along $\XTwoInt$.  
Let
\ifCLASSOPTIONtwocolumn 
\begin{align}
\dTwoInt &\equiv d_2' - \dTwoOrth \nonumber \\
&=\min \left\{ \!\!\!
\begin{tabular}{c}
$\dim \TxIntSpace{22}{12}$, \tabularnewline 
$(\dim \TxSpace{12} - \dim \RxSpace{12} )^+  + \dim \RxSubspace{12}{11}$ \tabularnewline
$\dim \RxSpace{22} - \dim \TxSubspace{22}{12}$
\end{tabular} 
\!\!\! \right\}.
\end{align}
\else
\begin{align}
\dTwoInt &\equiv d_2' - \dTwoOrth =\min \left\{ \!\!\!
\begin{tabular}{c}
$\dim \TxIntSpace{22}{12}$, \tabularnewline 
$(\dim \TxSpace{12} - \dim \RxSpace{12} )^+  + \dim \RxSubspace{12}{11},$
\tabularnewline
$(\dim \RxSpace{22} - \dim \TxSubspace{22}{12})^+$ 
\end{tabular} 
\!\!\! \right\}.
\label{eq:dTwoInt}
\end{align}
\fi
Now since $\XTwoInt\in \TxSpace{12}$, $ \ScatOp{12}\XTwoInt$ is nonzero in general,  $\XTwoInt$ will present interference to $R_1$. Therefore we must construct  $\XTwoInt$ such that it communicates $\dTwoInt$ symbols to $R_2$, without impeding $R_1$ from recovering the $d_1'$ symbols transmitted from $T_1$. Thus the construction of $\XTwoInt\in\TxSpace{12}$ will indeed depend on the structure of $\ScatOp{12}$. 

First consider the case where $\dim \TxSpace{12} \leq \dim \RxSpace{12}$. 
In this case Equation (\ref{eq:dTwoInt}), which gives the number of symbols that must be  transmitted along $\XTwoInt$, simplifies to $\dTwoInt = \min \{\dim \TxIntSpace{22}{12}, \dim \RxSubspace{12}{11}, (\dim \RxSpace{22} - \dim \TxSubspace{22}{12})^+\}$. Let
$$\left\{\SingVal{12}{k}, \SingLeft{12}{k}, \SingRight{12}{k}\right\}_{k=1}^{\infty}$$ be a singular system for $\ScatOp{12}$. From Property 3 of Lemma~\ref{lem:scatteringProperties}, we know that $\SingVal{12}{k}$ is zero for $k>\TxSpace{12}$ and nonzero for $k\leq \TxSpace{12}$. Note that 
$$ \left\{ \SingRight{12}{k} \right\}_{k=1}^{\dim \TxSpace{12}}$$ is an orthonormal basis for $\TxSpace{12}$. In the case of $\dim \TxSpace{12} \leq \dim \RxSpace{12}$ for which are constructing $\XTwoInt$
\begin{align}
\dTwoInt &= \min \{\dim \TxIntSpace{22}{12}, \dim \RxSubspace{12}{11}, (\dim \RxSpace{22} - \dim \TxSubspace{22}{12})^+\}, \quad \dim \TxSpace{12} \leq \dim \RxSpace{12} \\
& \leq \dim \TxIntSpace{22}{12} \\
&\leq \dim \TxSpace{12},
\end{align} 
so that there are at least as many $\SingRight{12}{k}$'s as there are symbols to transmit along $\XTwoInt$.  We construct $\XTwoInt$ as
\begin{equation}
\XTwoInt = \sum_{k=1}^{\dTwoInt} \TxSym{2}{k+\dim \dTwoOrth}\SingRight{12}{k}.
\end{equation}

Now we will consider the construction of $\XTwoInt$ for the other case where $\dim \TxSpace{12} > \dim \RxSpace{12}$.
In the $\dim \TxSpace{12} > \dim \RxSpace{12}$ case Equation (\ref{eq:dTwoInt}), which gives the number of symbols that must be  transmitted along $\XTwoInt$, simplifies to 
$$\dTwoInt = \min \left\{ \begin{tabular}{c}
$\dim \TxIntSpace{22}{12}$, \tabularnewline 
$(\dim \TxSpace{12} - \dim \RxSpace{12} )  + \dim \RxSubspace{12}{11},$
\tabularnewline
$(\dim \RxSpace{22} - \dim \TxSubspace{22}{12})^+$ 
\end{tabular} \right\},\qquad \dim \TxSpace{12} > \dim \RxSpace{12}.$$
Note that the signal that $R_1$ receives from $T_1$ will lie only in $\RxSpace{11}$. Thus if we can ensure that the signal from $T_2$ falls in the orthogonal space, $\RxSubspace{12}{11}$, then we have avoided interference. 
Let $\ScatOpRes{12}:\TxSpace{12}\rightarrow\RxSpace{12}$ be the restriction of $\ScatOp{12}:\TxSpace{2}\rightarrow\RxSpace{1}$ to domain $\TxSpace{12}$ and codomain $\RxSpace{12}$.\footnote{We consider tthe constriction, $\ScatOpRes{12}$, instead of $\ScatOp{12}$ so that the preimage under $\ScatOpRes{12}$ is subset of $\TxSpace{12}$, so that any functions within this preimage have not already been used in constructing $\XTwoOrth$. \evanNote{If I transmit in the preimage under $\ScatOp{12}$ then I may be doubling up. There may be a better way to do this, such as the way I argue the proof in the ISIT paper. }}
We can characterize the requirement that $Y_1(\tau)$ not be interfered over $\tau \in \IntR{11}$ as 
\begin{equation}
\label{eq:noInt}
\ScatOpRes{12}\XTwoInt \in \RxSubspace{12}{11},
\end{equation}
or equivalently 
\begin{equation}
\XTwoInt \in \TxTwoPreSpace,
\end{equation}
where
\begin{equation}
\TxTwoPreSpace \equiv \preim{\RxSubspace{12}{11}}{\ScatOpRes{12}} \subseteq \TxSpace{12},
\end{equation}
is the preimage of $\RxSubspace{12}{11}$ under $\ScatOpRes{12}$.
Thus any function in $\TxTwoPreSpace$ can be used for signaling to $R_2$ without interfering $X_1$ at $R_1$. The number of symbols that can be transmitted will thus depend on the dimension of this interference-free preimage.
%
Corollary~\ref{lem:preimSubset} in the appendix states that if 
$\Op{C}:\mathcal{X}\rightarrow \mathcal{Y}$ is a linear operator with closed range, and $\mathcal{S}$  is a subspace of the range of $\Op{C}$, $\mathcal{S} \subset R(\Op{C})$, then 
$\dim \preim{\mathcal{S}}{\Op{C}} = \dim N(\Op{C}) + \dim(\mathcal{S})$. Note that $R(\ScatOpRes{12})$ has finite dimension (namely 2$\min\{\LenTwoTx \IntT{12},\LenOneRx\IntR{12} \}<\infty$), and since any finite dimensional subspace of a normed space is closed,  $R(\ScatOpRes{12})$ is closed.
%
Further note that since we are considering the case where $\dim \TxSpace{12} > \dim \RxSpace{12}$, it is easy to see that $R(\ScatOpRes{12}) = \RxSpace{12}$, which implies  $\RxSubspace{12}{11}\subseteq R(\ScatOpRes{12})$, since $\RxSubspace{12}{11}\subseteq \RxSpace{12}$ by construction.
Thus the linear operator $\ScatOpRes{12}:\TxSpace{12}\rightarrow \RxSpace{12}$ and the subspace $\RxSubspace{12}{11}$ satisfy the conditions on operator $\Op{C}$ and subspace $\mathcal{S}$, respectively, in the hypothesis of Corollary~\ref{lem:preimSubset}. Thus we can apply Corollary~\ref{lem:preimSubset} to show that, when $\dim \TxSpace{12} > \dim \RxSpace{12}$, the dimension of $\TxTwoPreSpace$ is given by  
\begin{align} 
\dim\TxTwoPreSpace & =   \dim  N(\ScatOpRes{12}) + \dim\RxSubspace{1}{11} \label{eq:dimPre} \\
&= (\dim \TxSpace{12} - \dim \RxSpace{12} )  + \dim \RxSubspace{12}{11} \\ 
&\geq 
\min \left\{ \begin{tabular}{c}
$\dim \TxIntSpace{22}{12}$, \tabularnewline 
$(\dim \TxSpace{12} - \dim \RxSpace{12} )  + \dim \RxSubspace{12}{11},$
\tabularnewline
$(\dim \RxSpace{22} - \dim \TxSubspace{22}{12})^+$ 
\end{tabular} \right\} \\
&=  \dTwoInt, \quad \dim \TxSpace{12} > \dim \RxSpace{12}.
\end{align}
Therefore the dimension of $\TxTwoPreSpace$, the preimage of $\RxSubspace{12}{11}$ under $\ScatOpRes{12}$, is indeed large enough to allow $T_2$ to transmit the remaining $\dTwoInt$ symbols along the basis functions of $\dim\TxTwoPreSpace$. 
Let 
$$\ \left\{\TxBasisPre{i} \right\}_{i=1}^{\dim\TxTwoPreSpace}$$ 
be an orthonormal basis for $\TxTwoPreSpace$.  
Then we construct $\XTwoInt$ as
\begin{align}
\XTwoInt &= \sum_{k=1}^{\dTwoInt}\TxSym{2}{k+\dTwoOrth}\TxBasisPre{k}.
\end{align}
 
In summary, combining all cases we see that the wavevector transmitted by $T_2$ is  
\begin{align}
X_2 &= \XTwoOrth + \XTwoInt \\
&= \sum_{i=1}^{\dTwoOrth}\TxSym{2}{i} \TxBasisOrth{i}
	+ \sum_{k=1}^{\dTwoInt} \TxSym{2}{k+\dTwoOrth}
	\left( \SingRight{12}{k}1{\scriptstyle(\dim \TxSpace{12} \leq \dim \RxSpace{12})} 
	+ \TxBasisPre{k}1{\scriptstyle(\dim \TxSpace{12} > \dim \RxSpace{12})} \right)\\
&= \sum_{i=1}^{\dTwoOrth}\TxSym{2}{i} \TxBasisOrth{i}
	+ \sum_{i=1+\dTwoOrth}^{d_2'} \TxSym{2}{i}
	\left( \SingRight{12}{i-\dTwoOrth}1{\scriptstyle(\dim \TxSpace{12} \leq \dim \RxSpace{12})} 
	+ \TxBasisPre{i-\dTwoOrth}1{\scriptstyle(\dim \TxSpace{12} > \dim \RxSpace{12})} \right) \\
	& = \sum_{i=1}^{d_2'}\TxSym{2}{i}\genBasisTwo,\quad \text{where} \quad
	\genBasisTwo =  
\left\{
     \begin{array}{lr}
       \TxBasisOrth{i} & : i \leq \dim \dTwoOrth
       \\
       \SingRight{12}{i-\dim \dTwoOrth}  & : i > \dim \dTwoOrth,\  \dim \TxSpace{12} \leq \RxSpace{12}
       \\
        \TxBasisPre{i-\dim \dTwoOrth}  & : i > \dim \dTwoOrth,\ \dim \TxSpace{12} > \RxSpace{12}
     \end{array}.
 \right.
\end{align}

Now that we have constructed $X_1$, the uplink wavevector signal transmitted on the the uplink user, and $X_2$, the wavevector signal transmitted on the dowlink by the base station, we show how the base station receiver, $R_1$ and the downlink user $R_2$ process their received signals to detect the original information-bearing symbols.

\textbf{Processing at the base station receiver, $R_1$:}
We need to show that $R_1$ can obtain at least $d_1' = \dim \RxSpace{11}$ independent linear combinations of the $d_1'$ symbols transmitted from $T_1$, and that each of these linear combinations are corrupted only by noise, and not interference from $T_2$. 

In the case where $\dim \TxSpace{12} > \dim \RxSpace{12}$, $T_2$ constructed $X_2$ such that $\ScatOp{12}X_2$ is orthogonal to any function in $\RxSpace{11}$.
Therefore $R_1$ can eliminate interference from $T_2$ by simply projecting $Y_1$ onto $\RxSpace{11}$ to recover the $\dim \RxSpace{11}$ linear combinations it needs. We now formalize this projection onto $\RxSpace{11}$. Recall that the set of left-singular functions of $\ScatOp{11}$, $\{\SingLeft{11}{l} \}_{l=1}^{\dim \RxSpace{11}},$ form an orthonormal basis for $\RxSpace{11}$.
%
In the case where $\dim \TxSpace{12} > \dim \RxSpace{12}$, receiver $R_2$ constructs the set of complex scalars 
$$\left\{ \RxSym{1}{l} \right\}_{l=1}^{\dim \RxSpace{11}},\quad \RxSym{1}{l} = \langle   Y_1 , \SingLeft{11}{l} \rangle.$$  
One can check that result of each of these projections is
\begin{align}
\RxSym{1}{l} =  \SingVal{11}{l} \TxSym{1}{l} + \left\langle Z_1, \SingLeft{11}{l} \right\rangle,\quad l = 1,2,\dots, \dim \RxSpace{11}, 
\end{align}
and thus obtains each of the $d_1'=\dim \RxSpace{11}$ linear combinations of the intended symbols corrupted only by noise, as desired. Moreover, in this case the obtained linear combinations are already diagonalized, with the $l$th projection only containing a contribution from the $l$th desired symbol.

In the case where $\dim \TxSpace{12} \leq \dim \RxSpace{12}$, $\ScatOp{12}X_2$ in general will not be orthogonal to every function in $\RxSpace{11}$, and some slightly more sophisticated processing must be performed to decouple the interference from the signal of interest. 
First, $R_1$ can recover $\dim \RxSubspace{11}{12}$ interference-free linear combinations by projecting its received signal, $Y_1$, onto $\RxSubspace{11}{12}$.
Let 
$$\left\{\RxBasis{11\setminus12}{l}\right\}_{l=1}^{\dim \RxSubspace{11}{12}}$$
be an orthonormal basis for $\RxSubspace{11}{12}$. Receiver $R_1$ forms a set of complex scalars 
$$\left\{ \RxSym{1}{l} \right\}_{l=1}^{\dim \RxSubspace{11}{12}},\quad \RxSym{1}{l} = \langle Y_1 , \RxBasis{11\setminus12}{l} \rangle.$$ 
Note that each $\RxBasis{11\setminus12}{l}$ will be orthogonal to $\ScatOp{12}X_2$ for any $X_2$ since each $\RxBasis{11\setminus12}{l} \in \RxSubspace{11}{12}$, and $\ScatOp{12}X_2 \in \RxSpace{12}$ for any $X_2$, and $\RxSubspace{11}{12}$ is the orthogonal complement of $\RxSpace{12}$.
Therefore, each $\RxSym{1}{l}$ will be interference free, i.e., will be a linear combination of the symbols $\{\TxSym{1}{l}\}_{l=1}^{d_1'}$ plus noise, and will contain no contribution from the $\{\TxSym{2}{l}\}_{l=1}^{d_2'}$ symbols. 
One can check that these $\dim \RxSubspace{11}{12}$ projections result in
\begin{align}
\RxSym{1}{l} = \sum_{m=1}^{d_1'} \SingVal{11}{l}{\left\langle  \SingLeft{11}{m}, \RxBasis{11\setminus12}{l} \right\rangle} \TxSym{1}{m} + \left\langle Z_1, \RxBasis{11\setminus12}{l} \right\rangle,\quad l = 1,2,\dots, \dim \RxSubspace{11}{12}.
\end{align}
It remains to obtain
$d_1' - \dim \RxSubspace{11}{12}  
= \dim \RxSpace{11} - \dim \RxSubspace{11}{12}
= \dim \RxIntSpace{11}{12}$
more independent and interference-free linear combinations of $T_1$'s symbols so that $R_1$ can solve the system and recover the symbols. Receiver $R_1$ will obtain these linear combinations via a careful projection onto a subspace of $\RxSpace{12}$ (which is the orthogonal complement of $\RxSubspace{11}{12}$, the space onto which we have already projected $Y_1$ to obtain the first $\dim \RxSubspace{11}{12}$ linear combinations). 
Recall that the set of left-singular functions of $\ScatOp{12}$, $\{\SingLeft{12}{l} \}_{l=1}^{\dim \RxSpace{12}}$, form an orthonormal basis for $\RxSpace{12}$.
Receiver $R_1$ obtains the remaining $\RxIntSpace{11}{12}$ linear combinations 
by projecting $Y_1$ onto the last $\dim \RxIntSpace{11}{12}$ of these basis functions, forming $\{ \RxSym{1}{l}\}_{l=\dim \RxSubspace{11}{12}+1}^{\dim \RxSpace{11}}$ by computing 
%
%
\begin{align}
\quad \RxSym{1}{k + \dim \RxSubspace{11}{12}} &= \left\langle Y_1, \SingLeft{12}{\dim \RxSpace{12} - k} \right\rangle, \qquad  k = 0,1,\dots, \dim \RxIntSpace{11}{12}-1,\\
& = \left\langle \ScatOp{11}X_1 + \ScatOp{12}X_2 + Z_1, \SingLeft{12}{\dim \RxSpace{12} - k} \right\rangle
\\
& = \left\langle \ScatOp{11}X_1, \SingLeft{12}{\dim \RxSpace{12} - k} \right\rangle + \left\langle\ScatOp{12}X_2, \SingLeft{12}{\dim \RxSpace{12} - k} \right\rangle + \left\langle Z_1, \SingLeft{12}{\dim \RxSpace{12} - k} \right\rangle \label{eq:threeTerms}.
\end{align}
We compute the terms of Equation~(\ref{eq:threeTerms}) individually. First, the contribution of $T_1$'s transmit wavevector is 
\begin{align}
\left\langle 
\ScatOp{11}X_1,\ \SingLeft{12}{\dim \RxSpace{12} - k} \right\rangle
&= \left\langle 
\sum_{m=1}^{d_1' = \min(\dim \TxSpace{11}, \dim \RxSpace{11})} \SingVal{11}{m} \SingLeft{11}{m} \langle \SingRight{11}{m}, X_1  \rangle
,\ \SingLeft{12}{\dim \RxSpace{12} - k}  
\right\rangle \\
&= 
\left\langle 
\sum_{m=1}^{d_1'} \SingVal{11}{m} \SingLeft{11}{m} \left\langle \SingRight{11}{m}, \sum_{i=1}^{d_1'} \TxSym{1}{i} \SingRight{11}{i}  \right\rangle
,\ \SingLeft{12}{\dim \RxSpace{12} - k} \right\rangle \\
&=  
\left\langle 
\sum_{m=1}^{d_1'}  \SingVal{11}{m} \SingLeft{11}{m} \sum_{i=1}^{d_1'} \TxSym{1}{i} \left\langle \SingRight{11}{m},  \SingRight{11}{i}  \right\rangle
,\ \SingLeft{12}{\dim \RxSpace{12} - k} \right\rangle \\
&=  
\left\langle 
\sum_{m=1}^{d_1'}  \SingVal{11}{m} \SingLeft{11}{m} \sum_{i=1}^{d_1'}  \TxSym{1}{i} \delta_{mi}
,\ \SingLeft{12}{\dim \RxSpace{12} - k} \right\rangle \\
&=  
\left\langle 
\sum_{m=1}^{d_1'} \SingVal{11}{m} \SingLeft{11}{m} \TxSym{1}{m} 
,\ \SingLeft{12}{\dim \RxSpace{12} - k} \right\rangle \\
&=  
\sum_{m=1}^{d_1'}  \SingVal{11}{m}\left\langle \SingLeft{11}{m}
,\ \SingLeft{12}{\dim \RxSpace{12} - k} \right\rangle \TxSym{1}{m},\quad  k = 0,1,\dots, \dim \RxIntSpace{11}{12}-1. \label{eq:SigTerm}
\end{align}
Second, the contribution of $T_2$'s interfering wavevector is 
\begin{align}
\left\langle 
\ScatOp{12}X_2,\ \SingLeft{12}{\dim \RxSpace{12} - k} \right\rangle
&= \left\langle 
\ScatOp{12} (\XTwoOrth + \XTwoInt)
,\ \SingLeft{12}{\dim \RxSpace{12} - k}  
\right\rangle \\
&= \left\langle 
\ScatOp{12} \XTwoInt
,\ \SingLeft{12}{\dim \RxSpace{12} - k}  
\right\rangle \\
&= \left\langle 
\sum_{m=1}^{\infty} \SingVal{12}{m} \SingLeft{12}{m} \langle \SingRight{12}{m}, \XTwoInt  \rangle
,\ \SingLeft{12}{\dim \RxSpace{12} - k}  
\right\rangle \\
&= \left\langle 
\sum_{m=1}^{\min(\dim \TxSpace{12}, \dim \RxSpace{12})} \SingVal{12}{m} \SingLeft{12}{m} \langle \SingRight{12}{m}, \XTwoInt  \rangle
,\ \SingLeft{12}{\dim \RxSpace{12} - k}  
\right\rangle \\
&= \left\langle 
\sum_{m=1}^{\dim \TxSpace{12}} \SingVal{12}{m} \SingLeft{12}{m} \langle \SingRight{12}{m}, \XTwoInt  \rangle
,\ \SingLeft{12}{\dim \RxSpace{12} - k}  
\right\rangle \\
&= \left\langle 
\sum_{m=1}^{\dim \TxSpace{12}} \SingVal{12}{m} \SingLeft{12}{m} \left\langle \SingRight{12}{m}, \sum_{i=1}^{\dTwoInt} \TxSym{2}{i+\dim \dTwoOrth}\SingRight{12}{i}  \right\rangle
,\ \SingLeft{12}{\dim \RxSpace{12} - k}  
\right\rangle \\
&= \left\langle 
\sum_{i=1}^{\dTwoInt} \TxSym{2}{i+\dim \dTwoOrth} \sum_{m=1}^{\dim \TxSpace{12}}  \SingVal{12}{m} \SingLeft{12}{m}  \left\langle \SingRight{12}{m},  \SingRight{12}{i}  \right\rangle
,\ \SingLeft{12}{\dim \RxSpace{12} - k}  
\right\rangle \\
&= \left\langle 
\sum_{i=1}^{\dTwoInt} \TxSym{2}{i+\dim \dTwoOrth} \sum_{m=1}^{\dim \TxSpace{12}}  \SingVal{12}{m} \SingLeft{12}{m}  \delta_{im},\ \SingLeft{12}{\dim \RxSpace{12} - k}  
\right\rangle \\
&= \left\langle 
\sum_{i=1}^{\dTwoInt} \TxSym{2}{i+\dim \dTwoOrth}  \SingVal{12}{i} \SingLeft{12}{i},\ \SingLeft{12}{\dim \RxSpace{12} - k}  
\right\rangle \\
&= 
\sum_{i=1}^{\dTwoInt} \TxSym{2}{i+\dim \dTwoOrth}  \SingVal{12}{i}\left\langle  \SingLeft{12}{i},\ \SingLeft{12}{\dim \RxSpace{12} - k}  
\right\rangle,\quad  k = 0,1,\dots, \dim \RxIntSpace{11}{12}-1.  \\
&=
\sum_{i=1}^{\dTwoInt} \TxSym{2}{i+\dim \dTwoOrth}  \SingVal{12}{i}  \delta_{(i,\dim \RxSpace{12} - k)} 
,\quad  k = 0,1,\dots, \dim \RxIntSpace{11}{12}-1.  \\
&=0,\quad  k = 0,1,\dots, \dim \RxIntSpace{11}{12}-1, \label{eq:IntTerm}
\end{align}
where in the last step we have leveraged that when $\dim \TxSpace{12}\leq \dim \RxSpace{12} $, $\dTwoInt \leq \dim \RxSubspace{12}{11}$ (see Equation~\ref{eq:dTwoInt}), which means the largest value of $i$ in the summation, $\dTwoInt$, is smaller that the smallest value of $\dim \RxSpace{12} - k$ under consideration, $\dim \RxSpace{12} - \RxIntSpace{11}{12} + 1 = \dim \RxSubspace{12}{11} + 1$, so that $\delta_{(i,\dim \RxSpace{12} - k)}$ will never evaluate to one.
Substituting (\ref{eq:SigTerm}) and (\ref{eq:IntTerm}) back into (\ref{eq:threeTerms}) shows that the output symbols obtained by projecting $Y_1$ onto the last $\RxIntSpace{11}{12}$ functions of $\{\SingLeft{12}{l} \}_{l=1}^{\dim \RxSpace{12}}$ are
%
\begin{align}
\RxSym{1}{k + \dim \RxSubspace{11}{12}} = 
\sum_{m=1}^{d_1'}  \SingVal{11}{m}{\left\langle \SingLeft{11}{m} , \SingLeft{12}{\dim \RxSpace{12} - k} \right\rangle} \TxSym{1}{m} + 
\left\langle Z_1, \SingLeft{12}{{\dim \RxSpace{12} - k}} \right\rangle,\ \quad
k = 0,1,\dots, \dim \RxIntSpace{11}{12}-1.
\end{align}
%

Combining the processing in all cases, we see that receiver $R_1$ has formed a set of $d_1'$ complex scalars $\{ \RxSym{1}{l}\}_{l=1}^{d_1'}$, such that 
\begin{align}
\RxSym{1}{l} = \sum_{m=1}^{d_1'}\RxOneCoeffs{lm} \TxSym{1}{m} + \RxNoise{1}{l},\qquad l=1,2, \dots, d_1',
\end{align}
where 
\begin{equation}
\label{eq:coefs1}
\RxOneCoeffs{lm} =  
\left\{
     \begin{array}{lr}
     \delta_{lm} \SingVal{11}{l} 
     & : \dim \TxSpace{12} > \dim \RxSpace{12}
     \\
        \SingVal{11}{m}{\left\langle  \SingLeft{11}{m}, \RxBasis{11\setminus12}{l} \right\rangle} 
        & :\dim \TxSpace{12} \leq \dim \RxSpace{12} ,\  l \leq \dim \RxSubspace{11}{12}
       \\
        \SingVal{11}{m}{\left\langle \SingLeft{11}{m} , \SingLeft{12}{\dim \RxSpace{12} + \dim \RxSubspace{11}{12}-l} \right\rangle} 
        & :\dim \TxSpace{12} \leq \dim \RxSpace{12} ,\ l > \dim \RxSubspace{11}{12}
\end{array},
 \right.
\end{equation}
and
\begin{equation}
\label{eq:noise1}
\RxNoise{1}{l} =  
\left\{
     \begin{array}{lr}
     {\left\langle Z_1 , \SingLeft{11}{l} \right\rangle}  
     & : \dim \TxSpace{12} > \dim \RxSpace{12}
     \\
        {\left\langle  Z_1, \RxBasis{11\setminus12}{l} \right\rangle} 
        & :\dim \TxSpace{12} \leq \dim \RxSpace{12} ,\  l \leq \dim \RxSubspace{11}{12}
       \\
        {\left\langle Z_1 , \SingLeft{12}{\dim \RxSpace{12} + \dim \RxSubspace{11}{12}-l} \right\rangle} 
        & :\dim \TxSpace{12} \leq \dim \RxSpace{12} ,\ l > \dim \RxSubspace{11}{12}
\end{array}.
 \right.
\end{equation} 
Thus as desired, in all cases the base station receiver $R_1$ is able to obtain $d_1'$ interference-free linear combinations of the $d_1'$ symbols  by from the uplink user transmitter $T_1$. Now we move to the processing at the downlink user receiver.

\textbf{Processing at $R_2$}:
We wish to show that the downlink receiver, $R_2$, can recover the $d_2'$ symbols transmitted by the base station transmitter, $T_1$. 
Let $\{\SingVal{22}{k}, \SingLeft{22}{k}, \SingRight{22}{k}\}$ be a singular system for the operator $\ScatOp{22}$, and let \\ $r_{22}\equiv\min\left\{2\LenTwoTx |\IntT{22}|,2\LenTwoRx |\IntR{22}|\right\}$.
From Property 2 of Lemma~\ref{lem:scatteringProperties} we know that 
$\SingVal{22}{k}$ is zero for all $k > r_{22}$ and nonzero for $k \leq r_{22}$, so that
\begin{align} 	
Y_2 &= \ScatOp{22}X_2 + Z_2\\
&= \sum_{k=1}^{r_{22}} \SingVal{22}{k} \SingLeft{22}{k}  \langle \SingRight{22}{k}, X_2 \rangle + Z_2.
\end{align}
Receiver $R_2$ processes its received signal, $Y_2$, by projecting it onto each the first $d_2' \leq r_{22}$ left singular functions,\footnote{We could project onto all $r_{22}$ of the left singular functions which have nonzero singular values, but projecting onto the first $d_2'$ is sufficient to achieve the optimal spatial degrees-of-freedom.} forming a set of complex scalars $\left\{ \RxSym{2}{l} \right\}_{l=1}^{d_2'}$, where $\RxSym{2}{l} = \langle \SingLeft{22}{l} , Y_2 \rangle$. One can check that
\begin{align}
\RxSym{2}{l}  = \langle \SingLeft{22}{l} , Y_2 \rangle
	      = \sum_{m=1}^{d_2'} \RxTwoCoeffs{lm} \TxSym{2}{m} + \RxNoise{2}{l},
	      \qquad l = 1,\dots,d_2',
\end{align}
where 
\begin{equation}
\label{eq:coefs2}
\RxTwoCoeffs{lm} =  
\left\{
     \begin{array}{lr}
       \SingVal{22}{l}{\left\langle \SingRight{22}{l} , 
       \TxBasisOrth{m} \right\rangle} & : m \leq \dim \dTwoOrth
       \\
       \SingVal{22}{l}{\left\langle \SingRight{22}{l} , \SingRight{12}{m-\dim \dTwoOrth} \right\rangle} & : m > \dim \dTwoOrth,\  \dim \TxSpace{12} \leq \RxSpace{12}
       \\
       \SingVal{22}{l}{\left\langle \SingRight{22}{l} , \TxBasisPre{m-\dim \dTwoOrth} \right\rangle} & : m > \dim \dTwoOrth,\ \dim \TxSpace{12} > \RxSpace{12}
     \end{array}
 \right.
\end{equation}  
and
\begin{equation}
\label{eq:noise2}
\RxNoise{2}{l} = \langle \SingLeft{22}{l} , Z_2 \rangle. 
\end{equation}
%

\subsubsection{Reducing to parallel point-to-point vector channels}
The above processing at each transmitter and receiver has allowed the receivers $R_1$ and $R_2$ to recover the symbols 
\begin{align}
\RxSym{1}{l} &= \sum_{m=1}^{d_1'}\RxOneCoeffs{lm} \TxSym{1}{m} + \RxNoise{1}{l},\qquad l=1,2, \dots, d_1', \label{eq:processed1}\\
\RxSym{2}{l}  &= \sum_{m=1}^{d_2'} \RxTwoCoeffs{lm} \TxSym{2}{m} + \RxNoise{2}{l}, \qquad l = 1,\dots,d_2'. \label{eq:processed2}
\end{align}
respectively, where the linear combination coefficients, $\RxOneCoeffs{lm}$ and $\RxTwoCoeffs{lm}$, are given in (\ref{eq:coefs1}) and (\ref{eq:coefs2}), respectively  and the additive noise on each of the recovered symbols,  $\RxNoise{1}{l}$ and  $\RxNoise{2}{l}$, are given in (\ref{eq:noise1}) and (\ref{eq:noise2}), respectively.
%
We can rewrite (\ref{eq:processed1}-\ref{eq:processed2}) in matrix notation as 
\begin{align}
\RxVect{1} &= \Mat{1}\TxVect{1} + \NoiseVect{1},  \\
\RxVect{2} &= \Mat{2}\TxVect{2} + \NoiseVect{2},
\end{align}
where $\TxVect{1}$ and $\TxVect{2}$ are the $d_1'\times 1$ and $d_2' \times 1$ vectors of input symbols for transmitters $T_1$ and $T_2$, respectively, $\NoiseVect{1}$ and $\NoiseVect{2}$ are the $d_1'\times 1$ and $d_2' \times 1$ vectors of additive noise, respectively, and $\Mat{1}$ and $\Mat{2}$ are $d_1'\times d_1'$ and $d_2'\times d_2'$ square matrices whose elements are taken from $\RxOneCoeffs{lm}$ and $\RxTwoCoeffs{lm}$, respectively.   
The matrices $\Mat{1}$ and $\Mat{2}$ will be full rank for all but a measure-zero set of channel response kernels. Also, since each of the $\RxNoise{j}{l}$'s are linear combinations of Gaussian random variables, the the noise vectors, $\NoiseVect{1}$ and $\NoiseVect{2}$, are Gaussian distributed. Therefore the spatial processing has reduced the original channel to two parallel full-rank Gaussian vectors channels: the first a $d_1'\times d_1'$  channel and the second $d_2' \times d_2'$ channel, which are well known \cite{Teletar99MIMO} have $d_1'$ and $d_2'$ degrees-of-freedom respectively. Therefore the spatial degrees-of-freedom pair $(d_1', d_2')$ is indeed achievable.  
\end{IEEEproof}

	\begin{lem}
\label{lem:corners}
The degree-of-freedom pairs $(d_1',d_2')$ and $(d_1'',d_2'')$, are the corner points of $\D$, that is 
\begin{align}
(d_1',d_2') & =  \left(\dOneMax, \min\{\dTwoMax, \dSumMax - \dOneMax\} \right) \label{eq:corner1} \\
(d_1'',d_2'') & = \left(\min\{\dOneMax,\dSumMax - \dTwoMax\},  \dTwoMax\right)  \label{eq:corner2}.
\end{align}
\end{lem}

	\begin{IEEEproof}
%
Note that it is sufficient to prove only Equation (\ref{eq:corner1}), as Equation (\ref{eq:corner2}) follows by the symmetry of the expressions. 
It is easy to see that $d_1' = \min\left\{2\LenOneTx |\IntT{11}|, 2\LenOneRx |\IntR{11}|\right\} = \dOneMax$, but it is not so obvious that $d_2' = \min\{\dTwoMax, \dSumMax - \dOneMax\}$. 
However, one can verify that $d_2' = \min\{\dTwoMax, \dSumMax - \dOneMax\}$ by evaluating the left- and right-hand sides for all combinations of the conditions 
\begin{align}
\LenOneTx|\IntT{11}|  &\lesseqgtr  \LenOneRx|\IntR{11}|,\\
\LenTwoTx|\IntT{12}|  &\lesseqgtr  \LenOneRx|\IntR{12}|
\end{align}
and observing equality in each of the four cases. Table~\ref{tab:fourcases} shows the expressions to which $d_2'$ and $\min\{\dTwoMax, \dSumMax - \dOneMax\}$ both simplify in each of the four possible cases. 
\begin{table}[htdp]
\caption{Verifying that the corner points of inner and outer bounds coincide}
\begin{center}
\begin{tabular}{c|c}
Case & $d_2' = \min\{\dTwoMax, \dSumMax - \dOneMax\}$ \tabularnewline
\hline
\begin{tabular}{c}
$\scriptstyle \LenOneTx|\IntT{11}|  \geq  \LenOneRx|\IntR{11}|,$ \tabularnewline
$\scriptstyle \LenTwoTx|\IntT{12}|  \geq  \LenOneRx|\IntR{12}|$
\end{tabular} 
& 
$\min\{\dTwoMax,\ 2\LenTwoTx|\IntT{22}\cup\IntT{12}| - 2\LenOneRx|\IntR{11}\cap\IntR{12}| \}$
\tabularnewline \hline
\begin{tabular}{c}
$\scriptstyle \LenOneTx|\IntT{11}|  \geq  \LenOneRx|\IntR{11}|,$ \tabularnewline
$\scriptstyle \LenTwoTx|\IntT{12}|  <  \LenOneRx|\IntR{12}|$
\end{tabular} 
& 
$\min\{\dTwoMax,\ 2\LenTwoTx|\IntT{22}\setminus\IntT{12}|+ 2\LenOneRx |\IntR{12}\setminus\IntR{11}|\}$
\tabularnewline \hline
\begin{tabular}{c}
$\scriptstyle \LenOneTx|\IntT{11}|  <  \LenOneRx|\IntR{11}|,$ \tabularnewline
$\scriptstyle \LenTwoTx|\IntT{12}|  \geq  \LenOneRx|\IntR{12}|$
\end{tabular} 
& 
$\min\{\dTwoMax,\ 2\LenTwoTx|\IntT{22}\cup\IntT{12}| + 2\LenOneRx|\IntR{11}\setminus\IntR{12}| - 2\LenOneTx|\IntT{11}| \}$
\tabularnewline \hline
\begin{tabular}{c}
$\scriptstyle \LenOneTx|\IntT{11}|  <  \LenOneRx|\IntR{11}|,$ \tabularnewline
$\scriptstyle \LenTwoTx|\IntT{12}|  <  \LenOneRx|\IntR{12}|$
\end{tabular} 
& 
$\min\{\dTwoMax,\ 2\LenTwoTx|\IntT{22}\setminus\IntT{12}| + 2\LenOneRx|\IntR{11}\cup\IntR{12}| - 2\LenOneTx|\IntT{11}| \}$
\tabularnewline \hline
\end{tabular}
\end{center}
\label{tab:fourcases}
\end{table}%

%


\end{IEEEproof}
Lemmas~\ref{thm:achievePoints} and \ref{lem:corners} show that the corner points of $\D$, $(d_1',d_2')$ and $(d_1'',d_2'')$ are achievable. And thus all other points within $\D$ are achievable via time sharing between the schemes that achieve the corner points.

	\subsection{Converse}
	\label{sec:converse}
	To establish the converse part of Theorem~\ref{thm:mainResult}, we must show the region $\D$, which we have already shown is achievable, is also an outer bound on the degrees-of-freedom, i.e., we want to show that if an arbitrary degree-of-freedom pair $(d_1,d_2)$ is achievable, then $(d_1,d_2) \in \D$. 
It is easy to see that if $(d_1,d_2)$ is achievable, then the singe-user constraints on $\D$, given in (\ref{eq:d1Bound}) and (\ref{eq:d2Bound}), must be satisfied as the degrees-of-freedom for each flow cannot be more than the point-to-point degrees-of-freedom shown in \cite{TsePoon05DOF_EM}. 
Thus the only step remaining in the converse is to establish an outer bound on the the sum degrees-of-freedom which coincides with $\dSumMax$, the sum-degrees-of-freedom constraint on the achievable region, $\D$, given in (\ref{eq:dSumBound}). 
%
Thus to conclude the converse argument, we will now prove the following Genie-aided outer bound on the sum degrees-of-freedom which coincides with the sum-degrees-of-freedom constraint on the achievable region
\begin{lem}
\label{lem:sumBound}
\begin{align}
{d_1 + d_2} \leq  \ \dSumMax =
 2\LenTwoTx |\IntT{22} \setminus \IntT{12}| 
+ 2\LenOneRx |\IntR{11} \setminus \IntR{12}|  
+2 \max (\LenTwoTx |\IntT{12}|,  \LenOneRx |\IntR{12}|).
\label{eq:dSumBoundRefresh}
\end{align}
\end{lem}
\begin{IEEEproof}
We prove Lemmma~\ref{lem:sumBound} by way of a Genie that aids the transmitters and receivers by enlarging the scattering intervals and lengthening the antenna arrays in a way that can only enlarge the degrees-of-freedom region. Applying the point-to-point bounds to the Genie-aided system in a careful way then establishes the outer bound.
Assume an arbitrary scheme achieves the degrees-of-freedom pair $(d_1,d_2)$.
Thus receivers $R_1$ and $R_2$ can decode their corresponding messages with probability of error approaching zero.
We must show that the assumption of $(d_1,d_2)$ being achievable implies the constraint in Equation (\ref{eq:dSumBoundRefresh}).


Let a Genie expand both scattering intervals at $T_2$ into the union of the two scattering intervals, that is expand $\IntT{22}$ and $\IntT{12}$ to 
$$\IntTPrime{22} = \IntTPrime{12} =  \GenieIntT \equiv  \IntT{22} \cup \IntT{12}.$$ 
Likewise the Genie expands the scattering intervals at $R_1$ into their union, that is expand $\IntR{11}$ and $\IntR{12}$ to  
$$\IntRPrime{11} = \IntRPrime{12}  \equiv  \GenieIntR  = \IntR{11} \cup \IntR{12}.$$ 
%
%
%
%
The Genie's expansion of $\IntT{22}$ to $\GenieIntT$ can only enlarge the degrees-of-freedom region, as $T_2$ could simply not transmit in the added interval $\GenieIntT \setminus \IntT{22}$ (i.e. ignore the added dimensions for signaling to $R_2$) to obtain the original scenario. Likewise expanding $\IntR{11}$ to $\GenieIntR$ will only enlarge the degrees-of-freedom region as  $R_1$ can ignore the the portion of the wavevector received over $\GenieIntR\setminus \IntR{11}$ to obtain the original scenario. 
However, expanding the interference scattering clusters, $\IntT{12}$ and $\IntR{12}$, to $\GenieIntT$ and $\GenieIntR$, respectively, can indeed shrink the degrees-of-freedom region due to the additional interference causes by the added overlap with the signal-of-interest intervals $\IntT{22}$ and $\IntR{22}$, respectively. We need a final Genie manipulation to compensate for this added interference, so that the net Genie manipulation can only enlarge the degrees-of-freedom region. Therefore, in the next step we will have the Genie lengthen the arrays at $T_2$ and $R_1$ sufficiently  to allow any interference introduced by expanding $\IntT{12}$ and $\IntR{12}$, to $\GenieIntT$ and $\GenieIntR$, respectively, to be zero-forced without sacrificing any previously available degrees of freedom. 
Expansion of $\IntT{12}$ to $\GenieIntT \equiv \IntT{22} \cup \IntT{12}$ causes the dimension of the interference that $T_2$ presents to $R_1$ to increase by at most $2\LenTwoTx|\IntT{22} \setminus \IntT{12}|$. Therefore, let the Genie also lengthen $R_1$'s array from $2\LenOneRx$ to $2\LenOneRxPrime = 2\LenOneRx +  2\LenTwoTx \frac{|\IntT{22} \setminus \IntT{12}|} {|\IntR{11} \cup \IntR{12}|}$, so that the dimension of the total receive space at $R_1$, $\dim\RxSpace{1}$, is increased from 
$\dim\RxSpace{1} = 2\LenOneRx |\IntR{12} \cup \IntR{12}| $ to 
\begin{align}
\dim\RxSpacePrime{1} &= 2\LenOneRxPrime |\IntR{12} \cup \IntR{12}|\\
& =  \left(2\LenOneRx +  2\LenTwoTx \frac{|\IntT{22} \setminus \IntT{12}|} {|\IntR{11} \cup \IntR{12}|}\right) |\IntR{12} \cup \IntR{12}|\\
&= 2\LenOneRx |\IntR{12} \cup \IntR{12}| + 2\LenTwoTx|\IntT{22} \setminus \IntT{12}| \\
& = \dim\RxSpace{1} + 2\LenTwoTx|\IntT{22} \setminus \IntT{12}|
\label{eq:RxSpaceIncrease}.
\end{align}
We observe in (\ref{eq:RxSpaceIncrease}), that the Genie's lengthening of the $T_2$ array by $2\LenTwoTx \frac{|\IntT{22} \setminus \IntT{12}|} {|\IntR{11} \cup \IntR{12}|}$ has increased the dimension of $R_1$'s total receive signal space by $ 2\LenTwoTx|\IntT{22} \setminus \IntT{12}|$, which is the worst case increase in the dimension of the interference from $T_2$ due to expansion of $\IntT{12}$ to $ \IntT{22} \cup \IntT{12}$. 
Therefore the dimension of the subspace of $\RxSpacePrime{1}$ which is orthogonal to the interference from $T_2$ will be at least as large as in the original orthogonal space of $\RxSpace{1}$. Thus the combined expansion of $\IntT{12}$ to $\GenieIntT$ and lengthening of the $R_1$ array to $\LenOneRxPrime$ can only enlarge the degrees-of-freedom  region. 
Analogously, expansion of $\IntR{12}$ to $\GenieIntR \equiv \IntR{11}\cup\IntR{12}$ increases the dimension of $\RxSpace{12}$, the subspace of $R_1$'s receive space which is vulnerable to interference from $T_2$, by at most $2\LenOneRx |\IntR{11}\setminus\IntR{12}|$. Therefore let the Genie lengthen $T_2$'s array from $2\LenTwoTx$ to $2\LenTwoTxPrime = 2\LenTwoTx +  2\LenOneRx \frac{|\IntR{11} \setminus \IntR{12}|} {|\IntT{22} \cup \IntT{12}|}$, so that the dimension of the transmit space at $T_2$, $\dim\TxSpace{2}$, is increased from 
$\dim\TxSpace{2} = 2\LenTwoTx |\IntT{22} \cup \IntT{12}| $ to 
\begin{align}
\dim\TxSpacePrime{2} 
&= 2\LenTwoTxPrime |\IntT{22} \cup \IntT{12}|  \\
& = \left(2\LenTwoTx +  2\LenOneRx \frac{|\IntR{11} \setminus \IntR{12}|} {|\IntT{22} \cup \IntT{12}|} \right)|\IntT{22} \cup \IntT{12}|
\\
&= 2\LenTwoTx |\IntT{22} \cup \IntT{12}| + 2\LenOneRx |\IntR{11}\setminus\IntR{12}|
\\
&= \dim\TxSpace{2} + 2\LenOneRx |\IntR{11}\setminus\IntR{12}|\label{eq:TxSpaceIncrease}.
\end{align}
We see in (\ref{eq:TxSpaceIncrease}) that the Genie's lengthening of $T_2$'s array to $2\LenTwoTxPrime$ increases the dimension of $T_2$'s transmit signal space by $2\LenOneRx |\IntR{11}\setminus\IntR{12}|$, which is the worst case  increase in the dimension of the subspace of $R_1$'s receive subspace vulnerable to interference from $T_2$. Therefore $T_1$ can leverage these extra $2\LenOneRx |\IntR{11}\setminus\IntR{12}|$ dimensions to zero force to the subspace of $R_1$'s receive space that has become vulnerable to interference from $T_2$ due to the expansion $\IntR{12}$ to $\GenieIntR$. Thus the net effect of the Genie's expansion of $T_2$'s interference scattering interval, $\IntR{12}$, to $\GenieIntR$ and lengthening of the $T_2$ array to $2\LenTwoTxPrime$ can only enlarge the degrees-of-freedom region. 

\begin{figure}[htbp]
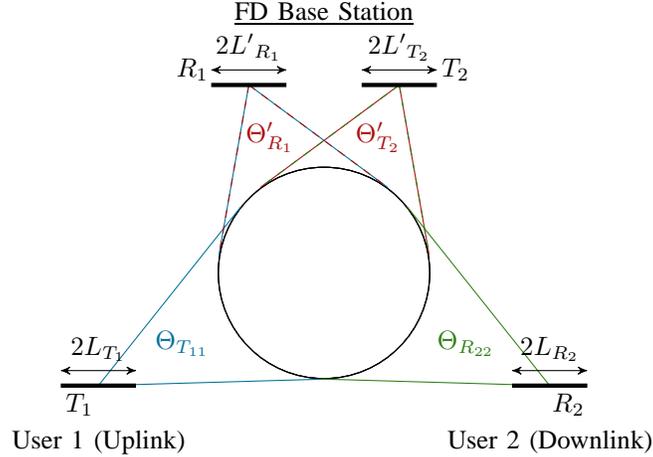

\begin{center}
\ifCLASSOPTIONtwocolumn \scalebox{0.9}{ \fi
\tikzstyle{block} = [draw,fill=KeynoteBlue!30,minimum size=2em]
\tikzstyle{bigBlock} = [draw,fill=KeynoteRed!50,minimum size=1.5em]
\tikzstyle{scatterer} = [draw,fill=gray!50,minimum size=2em]
\tikz[auto, decoration={
  markings,
  mark=at position 1.0 with {\arrow{>}}}
]{
\coordinate (T1) at (-3,0);
\coordinate (R1) at (-1,4);
\coordinate (T2) at (1,4);
\coordinate (R2) at (3,0);
%
\node [circle,draw] (S11) at (0,1.5) [minimum size=80pt] {};
\node [circle,draw] (S12) at (0,1.5) [minimum size=80pt] {};
\node [circle,draw] (S22) at (0,1.5) [minimum size=80pt] {};
\draw[KeynoteBlue] (T1) -- (tangent cs:node=S11,point={(T1)},solution=1);
\draw[KeynoteBlue] (T1) -- (tangent cs:node=S11,point={(T1)},solution=2);
\draw[KeynoteBlue] (R1) -- (tangent cs:node=S11,point={(R1)},solution=1);
\draw[KeynoteBlue] (R1) -- (tangent cs:node=S11,point={(R1)},solution=2);
\draw[KeynoteRed] (T2) -- (tangent cs:node=S12,point={(T2)},solution=1);
\draw[KeynoteRed] (T2) -- (tangent cs:node=S12,point={(T2)},solution=2);
\draw[KeynoteRed, dashed] (R1) -- (tangent cs:node=S12,point={(R1)},solution=1);
\draw[KeynoteRed, dashed] (R1) -- (tangent cs:node=S12,point={(R1)},solution=2);
\draw[KeynoteGreen, dashed] (T2) -- (tangent cs:node=S22,point={(T2)},solution=1);
\draw[KeynoteGreen, dashed] (T2) -- (tangent cs:node=S22,point={(T2)},solution=2);
\draw[KeynoteGreen] (R2) -- (tangent cs:node=S22,point={(R2)},solution=1);
\draw[KeynoteGreen] (R2) -- (tangent cs:node=S22,point={(R2)},solution=2);
\node[KeynoteBlue] at ($(S11.center)!60pt!(T1)$) {$\ElevT{11}$};
\node[KeynoteGreen] at ($(S22.center)!60pt!(R2)$) {$\ElevR{22}$};
%
\node[KeynoteRed] at ($(S12.center)!55pt!(T2)$) {$\GenieElevT$};
\node[KeynoteRed] at ($(S12.center)!55pt!(R1)$) {$\GenieElevR$};
%
%
\path (T1)+(-0.25,-0.25) node {$T_1$}
(T2)+(.75,.20) node {$T_2$}
(R1)+(-.75,.20) node {$R_1$}
(R2)+(+0.25,-0.25) node {$R_2$};
\draw[ultra thick] ($(R1)+(-.5,0)$)--($(R1)+(.5,0)$);
\draw[<->] ($(R1)+(-.5,.20)$)  to node {$2\LenOneRxPrime$} ($(R1)+(.5,.20)$);
\draw[ultra thick] ($(T2)+(-.5,0)$)--($(T2)+(.5,0)$);
\draw[<->] ($(T2)+(-.5,.20)$)  to node {$2\LenTwoTxPrime$} ($(T2)+(.5,.20)$);
\draw[ultra thick] ($(R2)+(-.5,0)$)--($(R2)+(.5,0)$);
\draw[<->] ($(R2)+(-.5,.20)$)  to node {$2\LenTwoRx$} ($(R2)+(.5,.20)$);
\draw[ultra thick] ($(T1)+(-.5,0)$)--($(T1)+(.5,0)$);
\draw[<->] ($(T1)+(-.5,.20)$)  to node {$2\LenOneTx$} ($(T1)+(.5,.20)$);
\node at ($(R1)!0.5!(T2)+(0,.95)$) {\underline{FD Base Station}};
\node at ($(T1)+(0,-0.75)$) {{User 1 (Uplink)}};
\node at ($(R2)+(0,-0.75)$) {{User 2 (Downlink)}};
}
\ifCLASSOPTIONtwocolumn } \fi
 
\caption{Genie-aided channel model
}
\label{fig:genireChannel}
\end{center}
\end{figure}

The Genie-aided channel is illustrated in Figure~\ref{fig:genireChannel}, which emphasizes the fact that the Genie has made the channel \emph{fully-coupled} in the sense that the signal-of-interest scattering and the interference scattering intervals are identical: any direction of departure from $T_2$ which scatters to $R_2$ also scatters to $R_1$, and any direction of arrival to $R_1$ which signal can be received from $T_1$ is a direction from which signal can be received from $T_2$.  %
%
Note that for the Genie-aided channel,
\begin{align}
\max(\dim\TxSpacePrime{2}, \dim\RxSpacePrime{1}) 
&= 
2\max(\LenTwoTxPrime|\GenieIntT|, \LenOneRxPrime|\GenieIntR|) \\
&= 
2\max \left\{ \begin{tabular}{c}
$\left(\LenTwoTx + \LenOneRx \frac{|\IntR{11}\setminus{\IntR{12}|}}{|\GenieIntT|}\right)|\GenieIntT|$, 
\tabularnewline 
$\left(\LenOneRx + \LenTwoTx \frac{|\IntT{22}\setminus{\IntT{12}|}}{|\GenieIntR|}\right)|\GenieIntR|$
\end{tabular} \right\} 
\\
&= 
2\max \left\{ \begin{tabular}{c}
$\LenTwoTx|\GenieIntT| + \LenOneRx |\IntR{11}\setminus\IntR{12}|$, 
\tabularnewline 
$\LenOneRx|\GenieIntR| + \LenTwoTx |\IntT{22}\setminus{\IntT{12}|}$
\end{tabular} \right\} 
\\
&= 
2\max \left\{ \begin{tabular}{c}
$\LenTwoTx|\IntT{22} \cup \IntT{12}| + \LenOneRx |\IntR{11}\setminus\IntR{12}|$, 
\tabularnewline 
$\LenOneRx|\IntR{11} \cup \IntR{12}| + \LenTwoTx |\IntT{22}\setminus{\IntT{12}|}$
\end{tabular} \right\} 
\\
&= 
2\max \left\{ \begin{tabular}{c}
$\LenTwoTx\left(|\IntT{12}| + |\IntT{22} \setminus \IntT{12}|\right) + \LenOneRx |\IntR{11}\setminus\IntR{12}|$, 
\tabularnewline 
$\LenOneRx\left(|\IntR{12}| + |\IntR{11} \setminus \IntR{12}|\right) + \LenTwoTx |\IntT{22}\setminus{\IntT{12}|}$
\end{tabular} \right\} 
\\
&= 
2\max \left\{ \begin{tabular}{c}
$\LenTwoTx|\IntT{12}| + \LenTwoTx|\IntT{22} \setminus \IntT{12}| + \LenOneRx |\IntR{11}\setminus\IntR{12}|$, 
\tabularnewline 
$\LenOneRx|\IntR{12}| + \LenOneRx|\IntR{11} \setminus \IntR{12}|+ \LenTwoTx |\IntT{22}\setminus{\IntT{12}|}$
\end{tabular} \right\} 
\\
&=  2\max (\LenTwoTx |\IntT{12}|,  \LenOneRx |\IntR{12}|)  + 2\LenTwoTx |\IntT{22} \setminus \IntT{12}| + 2\LenOneRx |\IntR{11} \setminus \IntR{12}|,   
\end{align}
which is the outer bound on sum degrees-of-freedom that we wish to prove. Thus if we can show that for the Genie-aided channel
\begin{equation}
\DoF \leq 2\max(\LenTwoTxPrime|\GenieIntT|, \LenOneRxPrime|\GenieIntR|) = \max(\dim\TxSpacePrime{2}, \dim\RxSpacePrime{1})\label{eq:toShow}
\end{equation}
then the converse is established. 
Because the Genie-aided channel is now fully coupled, it is similar to the continuous Hilbert space analog of the full-rank dicrete-atennas MIMO $Z$ interference channel. Thus the remaining steps in the converse argument are inspired by the techniques used in \cite{Ke12ZDoF,Jafar07MIMOInterferenceDoF,Jafar12DoFMIMOInterferenceRankDefficient} for outer bounding the degrees-of-freedom of the MIMO interference channel.  

Consider the case in which $\dim \TxSpacePrime{2} \leq \dim\RxSpacePrime{1}$. Since our Genie has enforced $\IntTPrime{22} = \IntTPrime{12}$ and we have assumed $\dim \TxSpacePrime{2}  \leq\dim\RxSpacePrime{1}$, receiver $R_1$ has access to the entire signal space of $T_2$, i.e., $T_2$ cannot zero force to $R_1$. Moreover, by our hypothesis that $(d_1,d_2)$ is achieved, $R_1$ can decode the message from $T_1$, and can thus reconstruct and subtract the signal received from $T_1$ from its received signal. 
%
Since $R_1$ has access to the entire signal-space of $T_2$, after removing the the signal from $T_1$ the only barrier to $R_1$ also decoding the message from $T_2$ is the receiver noise process. If it is not already the case, let a Genie lower the noise at receiver $R_1$ until $T_2$ has a better channel to $R_1$ than $R_2$ (this can only increase the capacity region since $R_1$ could always locally generate and add noise to obtain the original channel statistics). By hypothesis, $R_2$ can decode the message from $T_2$, and since $T_2$ has a better channel to $R_1$ than $R_2$, $R_1$ can also decode the message from $T_1$. 


Since $R_1$ can decode the messages from both $T_1$ and $T_2$, we can bound the degrees-of-freedom region of the Genie-aided channel by the corresponding point-to-point channel in which $T_1$ and $T_2$ cooperate to jointly communicate their messages to $R_1$, which has degrees-of-freedom $\min(\dim \TxSpacePrime{1} + \dim\TxSpacePrime{2},\ \dim\RxSpacePrime{1}$), which implies that
\begin{equation}
d_1 + d_2 \leq \dim \RxSpacePrime{1}, \quad \text{when } \dim \TxSpacePrime{2}  \leq\dim\RxSpacePrime{1}. \label{eq:outerBound1}
\end{equation}
Now consider the alternate case in which $\dim \TxSpacePrime{2}< \dim\RxSpacePrime{1}$. In this case we let a Genie increase the length of the $R_1$ array once more from $2\LenOneRxPrime$ to $2\LenOneRxPrimePrime =  2\LenTwoTxPrime \frac{|\GenieIntT|}{|\GenieIntR|} > 2\LenOneRxPrime$, so that the dimension of the receive signal space at $R_1$, which we now call $\RxSpacePrimePrime{1}$, is expanded to  
\begin{align}
\dim \RxSpacePrimePrime{1} &= 2\LenTwoRxPrime |\GenieIntR|
\\
&=\left( 2\LenTwoTxPrime \frac{|\GenieIntT|}{|\GenieIntR|} \right) |\GenieIntR|
\\
&= 2\LenTwoTxPrime |\GenieIntT| = \dim \TxSpacePrime{2}. 
\end{align}
Since $\dim \RxSpacePrimePrime{1} = \dim \TxSpacePrime{2}$ and $\IntTPrime{22} = \IntTPrime{12}$, $R_1$ again has access to the entire transmit signal space of $T_2$, we can use the same argument we leveraged above in the $\dim \TxSpacePrime{2}  \leq \dim\RxSpacePrime{1}$ case to show that 
\begin{equation}
d_1 + d_2 \leq \dim \RxSpacePrimePrime{1} = \dim \TxSpacePrime{2}, \quad \text{when } \dim \TxSpacePrime{2}  > \dim\RxSpacePrime{1}. \label{eq:outerBound2}
\end{equation}
Combining the bounds in (\ref{eq:outerBound1}) and (\ref{eq:outerBound2}) yields, 
\begin{align}
d_1 + d_2 &\leq  \max(\dim\TxSpacePrime{2}, \dim\RxSpacePrime{1}) \\
&= 2\max(\LenTwoTxPrime|\GenieIntT|, \LenOneRxPrime|\GenieIntR|) \\
&=  2\max (\LenTwoTx |\IntT{12}|,  \LenOneRx |\IntR{12}|)  + 2\LenTwoTx |\IntT{22} \setminus \IntT{12}| + 2\LenOneRx |\IntR{11} \setminus \IntR{12}|\label{eq:sumBoundShown}
\end{align} 
thus showing that the sum-degrees-of-freedom bound of Equation (\ref{eq:dSumBound}) in Theorem~\ref{thm:mainResult},  must hold for any achievable degree-of-freedom pair. 
\end{IEEEproof}

Combining Lemma~\ref{lem:sumBound} with  the trivial point-to-point bounds establishes that the region $\D$, given in Theorem~\ref{thm:mainResult} is an outer bound on any achievable degrees-of-freedom pair, thus establishing the converse part of Theorem~\ref{thm:mainResult}.

\section{Impact on Full-duplex Design}
\label{sec:impact}
We have characterized, $\D$, the degrees-of-freedom region achievable by a full-duplex base-station which uses spatial isolation to avoid self-interference while transmitting the uplink signal while simultaneously receiving. Now we wish to discuss how this result impacts the operation of full-duplex base stations. In particular, we aim to ascertain in what scenarios full-duplex with spatial isolation outperforms half-duplex, and are there scenarios in which full-duplex with spatial isolation achieves an ideal rectangular degrees-of-freedom regions (i.e. both the uplink flow and downlink flow achieving their respective point-to-point degrees-of-freedom).

To answer the above questions, we must first briefly characterize $\DHD$, the region of degrees-of-freedom pairs achievable via half-duplex mode, i.e. by time-division-duplex between uplink and downlink transmission. It is easy to see that the half-duplex achievable region is characterized by
\begin{align}
d_1 &\leq \alpha \min\left\{2\LenOneTx |\IntT{11}|, 2\LenOneRx |\IntR{11}|\right\},\\
d_2 & \leq (1-\alpha)\min\left\{2\LenTwoTx |\IntT{22}|, 2\LenTwoRx |\IntR{22}|\right\},
\end{align}
where $\alpha \in [0,1]$ is the time sharing parameter.  
Obviously $\DHD\subseteq \D$, but we are interested in contrasting the scenarios for which $\DHD\subset\D$, and full-duplex spatial isolation strictly outperforms half-duplex time division, and the scenarios for which $\DHD=\D$ and half-duplex can achieve the same performance as full-duplex. We will consider two particularly interesting cases: the fully spread environment, and the symmetric spread environment. 

\subsection{Overlapped Scattering Case}
Consider the worst case for full-duplex operation in which the self-interference backscattering intervals perfectly overlap the forward scattering intervals of the signals-of interest. By ``overlapped'' we mean that the directions of departure from the base station transmitter, $T_2$, that scatter to the intended downlink receiver, $R_2$, are identical to the directions of departure that backscatter to the base station receiver, $R_1$, as self-interference, so that $\IntT{11} = \IntT{12}$. Likewise the directions of arrival to the base station receiver, $R_1$, of the intended uplink signal from $T_1$ are identical to the directions of arrival of the backscattered self-interference from $T_2$, so that $\IntR{22} = \IntT{12}$. To reduce the number of variables in the degrees-of-freedom expressions, we assume each of the scattering intervals are of size $|\Int|$, so that 
$$
{|\IntT{11}| =  |\IntR{11}| = |\IntT{22}| = |\IntR{22}| = |\IntT{12}|= |\IntR{12}| \equiv |\Int|.}
$$
We further assume the base station arrays are of length $2\LenOneRx = 2\LenTwoTx = 2\LenBS$, and the user arrays are of equal length $2\LenOneTx = 2\LenTwoRx = 2\LenU$. In this case the full-duplex degrees-of-freedom region, $\D$, simplifies to 
\begin{align}
d_i \leq |\Int|\min\{2\LenBS,2\LenU\},i=1,2;\quad \label{eq:fullFD} 
d_1 + d_2 \leq 2\LenBS|\Int| 
\end{align}
while the half-duplex achievable region, $\DHD$ simplifies to
\begin{align}
d_1 + d_2 \leq |\Int|\min\{2\LenBS,2\LenU\}. \label{eq:fullHD} 
\end{align}
The following remark characterizes the scenarios for which full-duplex with spatial isolation beats half-duplex.
\begin{rem}
In the overlapped scattering case, $\DHD \subset \D$ when $2\LenBS > 2\LenU$, else $\DHD = \D$. 
\end{rem}
We see that full-duplex outperforms half-duplex only if the base station arrays are longer than the user arrays. This is because in the overlapped scattering case the only way to spatially isolate the self-interference is zero forcing, and zero forcing requires extra antenna resources at the base station. When $2\LenBS \leq 2\LenU$, the base station has no extra antenna resources it can leverage for zero forcing, and thus spatial isolation of the self-inference is no better than isolation via time division. However, when $2\LenBS > 2\LenU$ the base station transmitter can transmit  $(2\LenBS - 2\LenU)|\Int|$ zero-forced streams on the downlink without impeding the reception of the the full $2\LenU|\Int|$ streams on the uplink, enabling a sum-degrees-of-freedom gain of $(2\LenBS - 2\LenU)|\Int|$ over half-duplex. Indeed when the base station arrays are at least twice as long as the user arrays, the degrees-of-freedom region is rectangular, and both uplink and downlink achieve the ideal $2\LenU|\Int|$ degrees-of-freedom.


%
%

\subsection{Symmetric Spread}
The previous overlapped scattering case is worst case for full duplex operation. Let us now consider the more general case where the self-interference backscattering and the signal-of-interest forward scattering are not perfectly overlapped.  
This case illustrates the impact of the overlap of the scattering intervals on full-duplex performance. 
Once again, to reduce the number of variables, we will make following symmetry assumptions. Assume all the arrays in the network, the two arrays on the base station as well as the array on each of the user devices, are of the same length $2L$, that is 
$$2\LenOneTx =  2\LenOneRx = 2\LenTwoTx = 2\LenTwoRx \equiv 2L.$$
Also, assume that the size of the forward scattering intervals to/from the intended receiver/transmitter is the same for all arrays
$$|\IntT{11}| =  |\IntR{11}| = |\IntT{22}| = |\IntR{22}| \equiv |\IntFwd|,$$
and that the size of the backscattering interval is the same at the base station receiver as at the base station trasmitter 
$$ |\IntT{12}| = |\IntR{12}| \equiv |\IntBack|.$$
Finally assume the amount of overlap between the  backscattering and the forward scattering is the same at the base station transmitter as at the base station receiver so that
$$
|\IntT{22}\cap\IntT{12}| = |\IntR{11}\cap\IntR{12}| \equiv |\IntFwd \cap \IntBack| = |\IntFwd| - |\IntFwd \setminus \IntBack|.
$$

We call $\IntBack$ the \emph{backscatter interval} since it is the angle subtended at the base station by the back-scattering clusters, while we call  $\IntFwd$ the \emph{forward interval}, since it is the angle subtended by the clusters that scatter towards the intended transmitter/receiver. 
In this case, the full-duplex degree-of-freedom region, $\D$ simplifies to 
\begin{align}
d_i &\leq 2L|\IntFwd|,\ i=1,2 \label{eq:symInd}\\
d_1 + d_2 &\leq 2L(2 |\IntFwd\setminus\IntBack| + |\IntBack|) \label{eq:symSum}
\end{align}
while the half-duplex achievable region, $\DHD$ is 
\begin{align}
d_1 + d_2 &\leq 2L|\IntFwd|.
\end{align}

\begin{rem}
Comparing $\D$ and $\DHD$ above we see that in the case of symmetric scattering, $\DHD = \D$ if and only if $\IntFwd=\IntBack$,\footnote{We are neglecting the trivial case of $L=0$.} else $\DHD \subset \D$.
\end{rem}
%
Thus the full-duplex spatial isolation region is strictly larger than the half-duplex time-division region unless the forward interval and the backscattering interval are perfectly overlapped. The intuition is that when $\IntFwd=\IntBack$ the scattering interval is shared resource, just as is time, thus trading spatial resources is equivalent to trading time-slots. However, if $\IntFwd\neq\IntBack$, there is a portion of space exclusive to each user which can be leveraged to improve upon time division. Moreover, inspection of $\D$ above leads to the following remark. 

\begin{rem}
In the case of symmetric scattering, the full-duplex degree-of-freedom region is rectangular if and only if
\begin{equation}
|\IntBack\setminus\IntFwd| \geq |\IntFwd\cap\IntBack| \label{eq:sqCond}.
\end{equation}
\end{rem} 
The above remark can be verified by  comparing (\ref{eq:symInd}) and (\ref{eq:symSum}) observing that the sum-rate bound, (\ref{eq:symSum}),  is only active when 
\begin{equation}
2|\IntFwd\setminus\IntBack| + |\IntBack|  \geq 2|\IntFwd|. \label{eq:cond1}
\end{equation}
Straightforward set-algebraic manipulation of condition (\ref{eq:cond1}) shows that it is equivalent to (\ref{eq:sqCond}). 
The intuition is that because $\IntBack\setminus\IntFwd$ are the set directions in which the base station couples to itself but not to the users, the corresponding $2L|\IntBack\setminus\IntFwd|$ dimensions are useless for spatial multiplexing, and therefore ``free'' for zero forcing the self-interference, which has maximum dimension $2L|\IntFwd\cap\IntBack|$. Thus when $|\IntBack\setminus\IntFwd| \geq |\IntFwd\cap\IntBack|$, we can zero force any self-interference that is generated, without sacrificing any resource needed for spatial multiplexing to intended users. 

\begin{figure}[htbp]
\begin{center}
\begin{tikzpicture}
\begin{axis}[%
scale only axis,
width=2in,
height=2in,
xmin=0, xmax=2.5,
ymin=0, ymax=2.5,
xlabel={$d_1$},
ylabel={$d_2$},
ylabel near ticks,
xtick={.5,1,1.5,2, 2.5, 3},
xticklabels={$\frac{L}{2}$,$L$,$\frac{3L}{2}$,$2L$},
ytick={.5,1,1.5,2},
yticklabels={$\frac{L}{2}$,$L$,$\frac{3L}{2}$,$2L$},
legend entries={
{\footnotesize$\DHD$},
{\footnotesize$\D$: $|\IntFwd\cap\IntBack| = 1$},
{\footnotesize$\D$: $|\IntFwd\cap\IntBack| = 0.75$},
{\footnotesize$\D$: $|\IntFwd\cap\IntBack| \leq 0.5$}
},
legend style={ at={\ifCLASSOPTIONtwocolumn (0.5,1.1) \else (1.5,0.5) \fi}, anchor=center,nodes=right}]
%
%
\addplot [color=black, line width=2.0pt] coordinates{
	(2, 0)
	(0, 2)
};%
\addplot [color=green, line width=2.0pt, dotted] coordinates{
	(2, 0)
	(0, 2)
};%
\addplot [color=red, line width=2.0pt] coordinates{
	(2, 0)
	(2, 1)
	(1, 2)
	(0, 2)
};%
\addplot [color=blue, line width=2.0pt, dashed] coordinates{
	(2, 0)
	(2, 2)
	(0, 2)
};%
\end{axis}
\end{tikzpicture}
\caption{Symmetric-spread degree-of-freedom regions for different amounts of scattering overlap}
\label{fig:symRegions}
\end{center}
\end{figure}
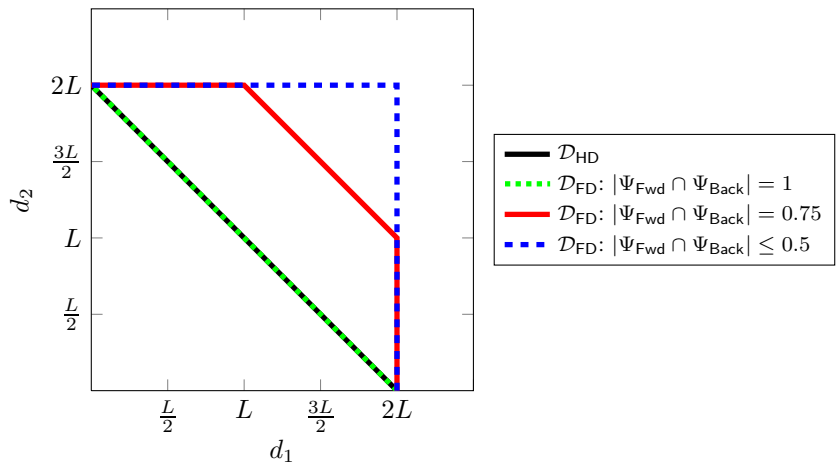

Consider a numerical example in which $|\IntFwd| = 1$ and $|\IntBack| = 1$, thus the overlap between the two, $|\IntFwd\cap\IntBack|$, can vary from zero to one. Figure~\ref{fig:symRegions} plots the half-duplex region, $\DHD$, and the full-duplex region, $\D$, for several different values  of overlap, $|\IntFwd\cap\IntBack|$. We see that when $\IntFwd=\IntBack$ so that $|\IntFwd\cap\IntBack|=1$, both $\DHD$ and $\D$ are the same triangular region. When $|\IntFwd\cap\IntBack|=0.75$, we get a rectangular region. Once $|\IntFwd\cap\IntBack|\leq 0.5$, $|\IntBack\setminus\IntFwd|$ becomes greater than 0.5, such that condition of (\ref{eq:sqCond}) is satisfied and the degree-of-freedom region becomes rectangular.

\section{Conclusion}
Full-duplex operation presents an opportunity for base stations to as much as double their spectral efficiency by both transmitting downlink signal and receiving uplink signal at the same time in the same band. The challenge to full-duplex operation is high-powered self-interference that is received both directly from the base station transmitter and backscattered from nearby objects. The receiver can be spatially isolated from the transmitter by leveraging multi-antenna beamforming to avoid self-interference, but such beamforming can also decrease the degrees-of-freedom of the intended uplink and downlink channels. We have leveraged a spatial antenna-theory-based channel model to analyze the spatial degrees-of-freedom available to a full-duplex base station. The analysis has shown the full-duplex operation can indeed outperform half-duplex operation when either (1) the base station arrays are large enough for for the base station to zero-force the backscattered self-interference or (2) the backscattering directions are not fully overlapped with the forward scattering directions, so that the base station can leverage the non-overlapped intervals for for interference free signaling to/from the intended users. 
\label{sec:conclusion}

\appendix
\section{Definitions and Lemmas}
\subsection{Definitions}
\label{subsec:defs}
The following definitions are standard in functional analysis, but are refreshed here for clarity. 

\begin{define}
Let $\mathcal{X}$ be a Hilbert space, the \emph{orthogonal complement} of $\mathcal{S} \subseteq \mathcal{X}$, denoted $\mathcal{S}^\perp$, is the subset 
$$ \mathcal{S}^\perp \equiv \{x \in \mathcal{X}: \langle x,u\rangle =0\ \forall\ u\in\mathcal{S} \}.$$
\end{define}

\begin{define}
Let $\mathcal{X}$ and $\mathcal{Y}$ be vector spaces (e.g., Hilbert spaces) and let $\Op{C}:\mathcal{X}\rightarrow \mathcal{Y}$ be a linear operator. Let $\mathcal{S} \subseteq \mathcal{Y}$ be a subspace of $\mathcal{Y}$.   
\begin{enumerate}[{(i)}]
\item The \emph{nullspace} of $\Op{C}$, denoted $N(\Op{C})$, is the subspace 
$$N(\Op{C}) \equiv \{x\in\mathcal{X}: \Op{C}x = 0 \}.$$  
\item The \emph{range} of $\Op{C}$, denoted $R(\Op{C})$, is the subspace 
$$R(\Op{C}) \equiv \{\Op{C}x: x\in \mathcal{X}\}.$$
\item The \emph{preimage} of $\mathcal{S}$ under $\Op{C}$, $\preim{\mathcal{S}}{\Op{C}}$, is the subspace (one can check that if $\mathcal{S}$ is a subspace then $\preim{\mathcal{S}}{\Op{C}}$ is a subspace also).  
$$\preim{\mathcal{S}}{\Op{C}} \equiv \{x \in \mathcal{X}: \Op{C}x \in \mathcal{S} \}.$$
\item The \emph{rank} of $\Op{C}$ is the dimension of the range of $\Op{C}$. A fundamental result in functional analysis is that the dimension of the range of $\Op{C}$ is also the dimension of the orthogonal complement of the nullspace of $\Op{C}$ (i.e. the coimage of $\Op{C}$) so that we can write
$$
\rank \Op{C} \equiv \dim R(\Op{C}) = \dim N(\Op{C})^\perp.
$$
\end{enumerate}
\end{define}

\subsection{Functional Analysis Lemmas}
\label{subsec:lems}

The following lemmas are general lemmas on compact operators on Hilbert spaces that will prove useful. 
We state without proof the following result from functional analysis (see Section 16.1 and 16.2 of \cite{Young88Hilbert} for proof).
\begin{lem}
\label{lem:SVD}
Let $\mathcal{X}$ and $\mathcal{Y}$ be Hilbert spaces and let $\Op{C}:\mathcal{X}\rightarrow \mathcal{Y}$ be a compact linear operator. There exists a \emph{singular system} $\{\sigma_k, {v}_k, {u}_k\}$, for $\Op{C}$ defined as follows. The set of functions $\{{u}_k\}$ form an orthonormal basis for $\overline{R(\Op{C})}$, the closure of the range of $\Op{C}$, and the set of functions $\{{v}_k\}$ form an orthonormal basis for ${N(\Op{C})}^\perp$, the coimage of $\Op{C}$. The set of positive real numbers $\sigma_k$, called the \emph{singular values} of $\Op{C}$, are the nonzero eigenvalues of $(\Op{C}^*\Op{C})$ arranged in decreasing order. The singular system diagonalizes $\Op{C}$ in the sense that for any $(\sigma_k, {v}_k, {u}_k) \in \{\sigma_k, {v}_k, {u}_k\}$, 
\begin{equation}
\Op{C} {v}_k = \sigma_k {u}_k.
\end{equation}
Moreover, the operation of $\Op{C}$ on any ${x}\in \mathcal{X}$ can be expanded as
\begin{equation}
\Op{C} {x} = \sum_k \sigma_k \langle  {x}, {v}_k \rangle {u}_k,
\end{equation}
which is called the singular value expansion of $\Op{C}{x}$.
\end{lem}

The next Lemma is a well-known characterization of the properties of the Moore-Penrose generalized inverse, and the conditions sufficient for its existence.    
\begin{lem} 
\label{lem:pseudo}
Let $\mathcal{X}$ and $\mathcal{Y}$ be Hilbert spaces and let $\Op{C}:\mathcal{X}\rightarrow \mathcal{Y}$ be a linear operator with closed range. There exists a unique linear operator $\Op{C}^+$, called the \emph{Moore-Penrose pseudoinverse} of $\Op{C}$, with the following properties:	\begin{enumerate}[{(i)}]
\item $\Op{C}^+ \Op{C} x = x\ \forall x \in N(\Op{C})^\perp$
\item $\Op{C} \Op{C}^+y = y\ \forall y\in R(\Op{C})$
\item $R(\Op{C}^+) = N(\Op{C})^\perp $
\item $N(\Op{C}^+) = R(\Op{C})^\perp $.
\end{enumerate}
\end{lem} 
\begin{IEEEproof}
See Definition 2.2 and Proposition 2.3 of \cite{Engl:InverseProblemsBook}. 
\end{IEEEproof}

\begin{lem}
\label{lem:preim}
Let $\mathcal{X}$ and $\mathcal{Y}$ be finite-dimensional Hilbert spaces and let $\Op{C}:\mathcal{X}\rightarrow \mathcal{Y}$ be a linear operator with closed range. Let $\mathcal{S} \subseteq \mathcal{Y}$ be a subspace of $\mathcal{Y}$. Then the dimension of the preimage of $\mathcal{S}$ under $\Op{C}$ is 
\begin{equation}
\dim \preim{\mathcal{S}}{\Op{C}} = \dim N(\Op{C}) + \dim(R(\Op{C})\cap \mathcal{S}).
\label{eq:preimLemma}
\end{equation}
\end{lem}

\begin{IEEEproof}
For notational convenience, let $\dPre \equiv \dim \preim{\mathcal{S}}{\Op{C}}$, $\dNull \equiv \dim N(\Op{C})$, and $\dInt \equiv \dim(R(\Op{C})\cap \mathcal{S})$. Thus we wish to show that $\dPre = \dNull + \dInt$. 
First note that $ N(\Op{C})  \subseteq \preim{\mathcal{S}}{\Op{C}} $, since $\mathcal{S}$ is a subspace and hence contains the zero vector, and the preimage of the zero vector under $\Op{C}$ is the nullspace of $\Op{C}$.  Denote the intersection between the preimage of $S$ under $\Op{C}$ and the orthogonal complement of the nullspace of $\Op{C}$ (i.e. the coimage) as
\begin{equation}
\mathcal{B} \equiv \preim{\mathcal{S}}{\Op{C}} \cap N(\Op{C})^\perp.
\end{equation}
Note that $\mathcal{B}$ is a subspace of $\mathcal{X}$ since the intersection of any collection of subspaces is itself a subspace (see Thm. 1 on p. 3 of \cite{LaxAnalysis}). 
Every $x\in\preim{\mathcal{S}}{\Op{C}}$ can be expressed as $x = w+u$ for some $w\in N(\Op{C})$ and $u \in \mathcal{B}$, and $\langle w , u \rangle = 0$ for any $w\in N(\Op{C})$ and $u \in \mathcal{B}$. 
Thus we can say that the preimage, $\preim{\mathcal{S}}{\Op{C}}$, is the \emph{orthogonal direct sum} of subspaces $N(\Op{C})$ and $\mathcal{B}$ (\cite{Young88Hilbert} Def. 4.26), a relationship we note we denote as 
\begin{equation}
\preim{\mathcal{S}}{\Op{C}} = N(\Op{C}) \oplus \mathcal{B}.
\end{equation}

Let $\{ a_i \}_{i=1}^{\dNull}$ be a basis for $N(\Op{C})$ and $\{ b_i \}_{i=1}^{\dB}$ be a basis for $\mathcal{B}$, where $\dNull = \dim N(\Op{C})$ and $\dB = \dim \mathcal{B}$. 
Construct the set $\{ e_i \}_{i=1}^{\dNull+\dB}$ according to 
\begin{equation}
\{ e_i \}_{i=1}^{\dNull} = \{ a_i \}_{i=1}^{\dNull},\qquad\{ e_i \}_{i=\dNull+1}^{\dNull+\dB} = \{ b_i \}_{i=1}^{\dB}.
\end{equation} 
We claim that $\{ e_i \}_{i=1}^{\dNull+\dB}$ forms a basis for $\preim{\mathcal{S}}{\Op{C}}$.
To check that $\{ e_i \}_{i=1}^{\dNull+\dB}$ is a basis for $\preim{\mathcal{S}}{\Op{C}}$, we must first show  $\{ e_i \}_{i=1}^{\dNull+\dB}$ spans $\preim{\mathcal{S}}{\Op{C}}$, and then show that the elements of $\{ e_i \}_{i=1}^{\dNull+\dB}$  are linearly independent. 
Consider an arbitrary $x\in \preim{\mathcal{S}}{\Op{C}}$. Since $\preim{\mathcal{S}}{\Op{C}} = N(\Op{C}) \oplus \mathcal{B}$, $x = w+u$ for some $w\in N(\Op{C})$ and $u \in \mathcal{B}$. Since by construction $\{ e_i \}_{i=1}^{\dNull}$ is a basis for $N(\Op{C})$ and $\{ e_i \}_{i=1+\dNull}^{\dNull+\dB}$ is a basis for $N(\Op{C})$, one can choose $\lambda_i$ such that that $w = \sum_{i=1}^{\dNull} \lambda_i e_i$ and $u = \sum_{i=1+\dNull}^{\dNull+\dB} \lambda_i e_i$. Thus 
\begin{equation}
x  = w + v = \sum_{i=1}^{\dNull} \lambda_i e_i + \sum_{i=1+\dNull}^{\dNull+\dB} \lambda_i e_i =  \sum_{i=1}^{\dNull+\dB} \lambda_i e_i
\end{equation} for some $\lambda_i$. Thus $\{ e_i \}_{i=1}^{\dNull+\dB}$ spans $\preim{\mathcal{S}}{\Op{C}}$. Now let us show linear independence: that $\sum_{i=1}^{\dNull+\dB} \lambda_i e_i = 0$ if and only if $\lambda_i = 0$ for all $i\in\{1,2,\dots,\dNull+\dB\}$. The ``if'' part is trivial, thus it remains to show that $\sum_{i=1}^{\dNull+\dB} \lambda_i e_i = 0$ implies $\lambda_i = 0\ \forall i$. The condition $\sum_{i=1}^{\dNull+\dB} \lambda_i e_i = 0$ implies 
\begin{align}
\label{eq:independence}
\sum_{i=1}^{\dNull} \lambda_i e_i = -\sum_{i=\dNull+1}^{\dNull+\dB} \lambda_i e_i,
\end{align}
which implies $w=-u$ for some $w\in N(\Op{C})$ and $u \in \mathcal{B}$. Every element of $N(\Op{C})$ is orthogonal to every element of $\mathcal{B}$ by construction, hence the only way Equation (\ref{eq:independence}) can be satisfied is if $w=u=0$, that is if both sides of Equation (\ref{eq:independence}) are zero, implying $\lambda_i = 0$ for all $i\in\{1,2,\dots,\dNull+\dB\}$ as desired. %
Thus we have shown $\{ e_i \}_{i=1}^{\dNull+\dB}$ is a basis for $\preim{\mathcal{S}}{\Op{C}}$, and hence 
\begin{equation}
\dPre = \dNull+\dB
\label{eq:dimCount}.
\end{equation}

Consider the set $\{\Op{C} e_i\}_{i=1+\dNull}^{\dNull+\dB}$.
By the definition of range, each element of the set $\{\Op{C} e_i\}_{i=1+\dNull}^{\dNull+\dB}$ is in $R(\Op{C})$, and since by construction each $e_i$ is in $\preim{\mathcal{S}}{\Op{C}}$, each element of $\{\Op{C} e_i\}_{i=1+\dNull}^{\dNull+\dB}$ is also in $\mathcal{S}$. 
We therefore have that 
\begin{equation}
\spanof \{\Op{C} e_i\}_{i=1+\dNull}^{\dNull+\dB} \subseteq R(\Op{C}) \cap \mathcal{S},
\end{equation}
and since there are $\dB$ elements in $\{\Op{C} e_i\}_{i=1+\dNull}^{\dNull+\dB}$, it must be that
\begin{equation}
\dB \leq \dInt.
\end{equation}
Substituting the above inequality into Equation (\ref{eq:dimCount}) gives 
\begin{equation}
\dPre \leq \dNull + \dInt.
\end{equation} 

To complete the proof we must show that $\dPre \geq \dNull + \dInt$. 
Let $\{s_i\}_{i=1}^{\dInt}$ be a basis for $R(\Op{C}) \cap \mathcal{S}$.
By assumption $R(\Op{C})$ is closed, thus we have by Lemma~\ref{lem:pseudo} that the Moore-Penrose pseudoinverse, $\Op{C}^+$, exists, and satisfies the properties listed in Lemma~\ref{lem:pseudo}.
Consider the set $\{\Op{C}^+ s_i\}_{i=1}^{\dInt}$. We claim that 
\begin{equation}
\label{eq:claim2}
\spanof \{\Op{C}^+ s_i\}_{i=1}^{\dInt} \subseteq N(\Op{C})^\perp \cap \preim{\mathcal{S}}{\Op{C}} \equiv  \mathcal{B} .
\end{equation}
By property (iv) in Lemma~\ref{lem:pseudo}, we have that $\Op{C}^+ s_i\in N(\Op{C})^\perp$ for each $\Op{C}^+ s_i\in \{\Op{C}^+ s_i\}_{i=1}^{\dInt}$.
Since $s_i \in R(\Op{C})$, we have that $\Op{C} (\Op{C}^+ s_i) = s_i$ by property (ii) of the pseudoinverse, and  since $s_i \in \mathcal{S}$, we have that $\Op{C} \Op{C}^+ s_i = s_i \in \mathcal{S}$ for each $\Op{C}^+ s_i\in \{\Op{C}^+ s_i\}_{i=1}^{\dInt}$. 
Thus each element of $\{\Op{C}^+ s_i\}_{i=1}^{\dInt}$ is also in $\preim{\mathcal{S}}{\Op{C}}$, the preimage of $\mathcal{S}$ under $\Op{C}$. Thus we have that each element of $ \{\Op{C}^+ s_i\}_{i=1}^{\dInt}$ is in $N(\Op{C})^\perp \cap \preim{\mathcal{S}}{\Op{C}}$ which justifies the claim of Equation (\ref{eq:claim2}). Now equation Equation (\ref{eq:claim2}) implies that 
\begin{equation}
\dInt \leq \dB
\end{equation}
Substituting the above inequality into Equation (\ref{eq:dimCount}) gives 
\begin{equation}
\dPre \geq \dNull + \dInt,
\end{equation} 
concluding the proof. 
\end{IEEEproof}

\begin{cor}
\label{lem:preimSubset}
Let $\mathcal{X}$ and $\mathcal{Y}$ be finite-dimensional Hilbert spaces and let $\Op{C}:\mathcal{X}\rightarrow \mathcal{Y}$ be a linear operator with closed range. Let $\mathcal{S} \subseteq R(\Op{C})\subseteq  \mathcal{Y}$ be a subspace of the range of $\Op{C}$. Then the dimension of the preimage of $\mathcal{S}$ under $\Op{C}$ is 
\begin{equation}
\dim \preim{\mathcal{S}}{\Op{C}} = \dim N(\Op{C}) + \dim(\mathcal{S}).
\end{equation}
\end{cor}

\begin{IEEEproof}
The proof follows trivially from Lemma~\ref{lem:preim} by noting that since $\mathcal{S} \subseteq  R(\Op{C})$, $R(\Op{C})\cap \mathcal{S} = \mathcal{S}$, which we substitute into equation~\ref{eq:preimLemma} to obtain the corollary. 
\end{IEEEproof}

\ifCLASSOPTIONcaptionsoff
  \newpage
\fi

\bibliographystyle{IEEEtran}
\bibliography{IEEEabrv,/Users/evaneverett/Dropbox/Wireless_Literature/Bibliography/Research}

\begin{thebibliography}{10}
\providecommand{\url}[1]{#1}
\csname url@samestyle\endcsname
\providecommand{\newblock}{\relax}
\providecommand{\bibinfo}[2]{#2}
\providecommand{\BIBentrySTDinterwordspacing}{\spaceskip=0pt\relax}
\providecommand{\BIBentryALTinterwordstretchfactor}{4}
\providecommand{\BIBentryALTinterwordspacing}{\spaceskip=\fontdimen2\font plus
\BIBentryALTinterwordstretchfactor\fontdimen3\font minus
  \fontdimen4\font\relax}
\providecommand{\BIBforeignlanguage}[2]{{%
\expandafter\ifx\csname l@#1\endcsname\relax
\typeout{** WARNING: IEEEtran.bst: No hyphenation pattern has been}%
\typeout{** loaded for the language `#1'. Using the pattern for}%
\typeout{** the default language instead.}%
\else
\language=\csname l@#1\endcsname
\fi
#2}}
\providecommand{\BIBdecl}{\relax}
\BIBdecl

\bibitem{Bliss07SimultTX_RX}
D.~W. Bliss, P.~A. Parker, and A.~R. Margetts, ``Simultaneous transmission and
  reception for improved wireless network performance,'' in \emph{Proceedings
  of the 2007 IEEE/SP 14th Workshop on Statistical Signal Processing}, 2007,
  pp. 478--482.

\bibitem{Khandani2010FDPatent}
A.~K. Khandani, ``Methods for spatial multiplexing of wireless two-way
  channels,'' US Patent US 7\,817\,641, October 19, 2010.

\bibitem{Radunovic:2009aa}
B.~Radunovic, D.~Gunawardena, P.~Key, A.~P.~N. Singh, V.~Balan, and G.~Dejean,
  ``Rethinking indoor wireless: Low power, low frequency, full duplex,''
  Microsoft Technical Report, Tech. Rep., 2009.

\bibitem{Duarte10FullDuplex}
M.~Duarte and A.~Sabharwal, ``Full-duplex wireless communications using
  off-the-shelf radios: Feasibility and first results,'' in \emph{Proc. 2010
  Asilomar Conference on Signals and Systems}, 2010.

\bibitem{Choi10FullDuplex}
J.~I. Choi, M.~Jain, K.~Srinivasan, P.~Levis, and S.~Katti, ``Achieving single
  channel, full duplex wireless communication,'' in \emph{MobiCom 2010}.\hskip
  1em plus 0.5em minus 0.4em\relax 

\bibitem{Jain2011RealTimeFD}
\BIBentryALTinterwordspacing
M.~Jain, J.~I. Choi, T.~Kim, D.~Bharadia, S.~Seth, K.~Srinivasan, P.~Levis,
  S.~Katti, and P.~Sinha, ``Practical, real-time, full duplex wireless,'' in
  \emph{Proceedings of the 17th annual international conference on Mobile
  computing and networking}, ser. MobiCom '11.\hskip 1em plus 0.5em minus
  0.4em\relax New York, NY, USA: ACM, 2011, pp. 301--312. [Online]. Available:
  \url{http://doi.acm.org/10.1145/2030613.2030647}
\BIBentrySTDinterwordspacing

\bibitem{Duarte11FullDuplex}
M.~Duarte, C.~Dick, and A.~Sabharwal, ``Experiment-driven characterization of
  full-duplex wireless systems,'' \emph{Wireless Communications, IEEE
  Transactions on}, vol.~11, no.~12, pp. 4296--4307, 2012.

\bibitem{Sahai11FullDuplex}
A.~Sahai, G.~Patel, and A.~Sabharwal, ``Pushing the limits of full-duplex:
  Design and real-time implementation,'' Rice Univeristy, Tech. Rep.
  {TREE1104}, 2011.

\bibitem{Khojastepour11AntennaCancellation}
\BIBentryALTinterwordspacing
M.~A. Khojastepour, K.~Sundaresan, S.~Rangarajan, X.~Zhang, and S.~Barghi,
  ``The case for antenna cancellation for scalable full-duplex wireless
  communications,'' in \emph{Proceedings of the 10th ACM Workshop on Hot Topics
  in Networks}, ser. HotNets-X.\hskip 1em plus 0.5em minus 0.4em\relax New
  York, NY, USA: ACM, 2011, pp. 17:1--17:6. [Online]. Available:
  \url{http://doi.acm.org/10.1145/2070562.2070579}
\BIBentrySTDinterwordspacing

\bibitem{Aryafar12MIDU}
\BIBentryALTinterwordspacing
E.~Aryafar, M.~A. Khojastepour, K.~Sundaresan, S.~Rangarajan, and M.~Chiang,
  ``Midu: enabling mimo full duplex,'' in \emph{Proceedings of the 18th annual
  international conference on Mobile computing and networking}, ser. Mobicom
  '12.\hskip 1em plus 0.5em minus 0.4em\relax New York, NY, USA: ACM, 2012, pp.
  257--268. [Online]. Available:
  \url{http://doi.acm.org/10.1145/2348543.2348576}
\BIBentrySTDinterwordspacing

\bibitem{Duarte2012FullDuplexWiFi}
M.~Duarte, A.~Sabharwal, V.~Aggarwal, R.~Jana, K.~Ramakrishnan, C.~Rice, and
  N.~Shankaranarayanan, ``Design and characterization of a full-duplex
  multiantenna system for wifi networks,'' \emph{Vehicular Tech., IEEE Trans.
  on}, vol.~63, no.~3, pp. 1160--1177, 2014.

\bibitem{Duarte12Thesis}
\BIBentryALTinterwordspacing
M.~Duarte, ``Full-duplex wireless: Design, implementation and
  characterization,'' Ph.D. dissertation, Rice University, April 2012.
  [Online]. Available: \url{http://warp.rice.edu/trac/wiki/DuartePhDThesis}
\BIBentrySTDinterwordspacing

\bibitem{Day12FDMIMO}
B.~Day, A.~Margetts, D.~Bliss, and P.~Schniter, ``Full-duplex bidirectional
  {MIMO}: Achievable rates under limited dynamic range,'' \emph{Signal
  Processing, IEEE Trans. on}, vol.~60, no.~7, pp. 3702 --3713, jul. 2012.

\bibitem{Day12FDRelay}
------, ``Full-duplex {MIMO} relaying: Achievable rates under limited dynamic
  range,'' \emph{Selected Areas in Communications, IEEE Journal on}, vol.~30,
  no.~8, pp. 1541 --1553, september 2012.

\bibitem{Everett11Asilomar}
E.~Everett, M.~Duarte, C.~Dick, and A.~Sabharwal, ``Empowering full-duplex
  wireless communication by exploiting directional diversity,'' in
  \emph{Asilomar Conference on Signals, Systems and Computers}, October 2011.

\bibitem{Everett2013PassiveSuppressionFD}
E.~Everett, A.~Sahai, and A.~Sabharwal, ``Passive self-interference suppression
  for full-duplex infrastructure nodes,'' \emph{Wireless Comm., IEEE Trans.
  on}, vol.~13, no.~2, pp. 680--694, Feb. 2014.

\bibitem{Sahai13PhaseNoise}
A.~Sahai, G.~Patel, C.~Dick, and A.~Sabharwal, ``On the impact of phase noise
  on active cancelation in wireless full-duplex,'' \emph{Vehicular Technology,
  IEEE Transactions on}, vol.~62, no.~9, pp. 4494--4510, 2013.

\bibitem{FullDuplexTutorial2014}
A.~Sabharwal, P.~Schniter, D.~Guo, D.~W. Bliss, S.~Rangarajan, and R.~Wichman,
  ``In-band full-duplex wireless: Challenges and opportunities,'' \emph{CoRR},
  vol. abs/1311.0456, 2013.

\bibitem{TsePoon05DOF_EM}
A.~Poon, R.~Brodersen, and D.~Tse, ``Degrees of freedom in multiple-antenna
  channels: a signal space approach,'' \emph{Information Theory, IEEE
  Transactions on}, vol.~51, no.~2, pp. 523 -- 536, feb. 2005.

\bibitem{TsePoon06EmagInfoTheory}
A.~Poon, D.~Tse, and R.~Brodersen, ``Impact of scattering on the capacity,
  diversity, and propagation range of multiple-antenna channels,''
  \emph{Information Theory, IEEE Transactions on}, vol.~52, no.~3, pp. 1087
  --1100, march 2006.

\bibitem{TsePoon11DOFPolarization}
A.~Poon and D.~Tse, ``Degree-of-freedom gain from using polarimetric antenna
  elements,'' \emph{Information Theory, IEEE Transactions on}, vol.~57, no.~9,
  pp. 5695 --5709, sept. 2011.

\bibitem{Poon03IndoorCharacterization}
A.~Poon and M.~Ho, ``Indoor multiple-antenna channel characterization from 2 to
  8 ghz,'' in \emph{Communications, 2003. ICC '03. IEEE International
  Conference on}, vol.~5, 2003, pp. 3519--3523 vol.5.

\bibitem{Spencer2000ClusteredChannels}
Q.~Spencer, B.~Jeffs, M.~Jensen, and A.~Swindlehurst, ``Modeling the
  statistical time and angle of arrival characteristics of an indoor multipath
  channel,'' \emph{Selected Areas in Communications, IEEE Journal on}, vol.~18,
  no.~3, pp. 347--360, 2000.

\bibitem{CramerUWBChannels}
R.~J.-M. Cramer, ``An evaluation of ultra-wideband propagation channels,''
  Ph.D. dissertation, University of Southern California, 2000.

\bibitem{Heddergott00NLoS_Channels}
R.~Heddergott and P.~Truffer, ``Statistical characteristics of indoor radio
  propagation in nlos scenarios,'' EUROPEAN COOPERATION IN THE FIELD OF
  SCIENTIFIC AND TECHNICAL RESEARCH, Valencia, Spain, Tech. Rep. COST 259
  TD(00) 024, 2000.

\bibitem{Young88Hilbert}
N.~Young, \emph{An Introduction to Hilbert Space}.\hskip 1em plus 0.5em minus
  0.4em\relax Cambridge University Press, 1988.

\bibitem{Engl:InverseProblemsBook}
H.~W. Engl, M.~Hanke, and A.~Neubauer, \emph{Regularization of Inverse
  Problems}.\hskip 1em plus 0.5em minus 0.4em\relax Kluwer Academic Publishers,
  1996.

\bibitem{Teletar99MIMO}
E.~Telatar, ``Capacity of multi-antenna gaussian channels,'' \emph{European
  Transactions on Telecommunications}, vol.~10, no.~6, pp. 585--595, 1999.

\bibitem{Ke12ZDoF}
L.~Ke and Z.~Wang, ``Degrees of freedom regions of two-user mimo z and full
  interference channels: The benefit of reconfigurable antennas,''
  \emph{Information Theory, IEEE Transactions on}, vol.~58, no.~6, pp.
  3766--3779, 2012.

\bibitem{Jafar07MIMOInterferenceDoF}
S.~Jafar and M.~Fakhereddin, ``Degrees of freedom for the mimo interference
  channel,'' \emph{Information Theory, IEEE Transactions on}, vol.~53, no.~7,
  pp. 2637--2642, 2007.

\bibitem{Jafar12DoFMIMOInterferenceRankDefficient}
S.~Krishnamurthy and S.~Jafar, ``Degrees of freedom of 2-user and 3-user
  rank-deficient mimo interference channels,'' in \emph{Global Communications
  Conference (GLOBECOM), 2012 IEEE}, Dec 2012, pp. 2462--2467.

\bibitem{LaxAnalysis}
P.~D. Lax, \emph{Funtional Analysis}.\hskip 1em plus 0.5em minus 0.4em\relax
  John Wiley and Sons Inc., 2002.

\end{thebibliography}
\end{document}